\documentclass[fleqn,usenatbib]{mnras}

\usepackage{newtxtext,newtxmath}

\usepackage[T1]{fontenc}
\usepackage{ae,aecompl}

\usepackage{amsfonts}
\usepackage{commath}
\usepackage{xfrac}
\usepackage{subfigure}
\usepackage{booktabs}
\usepackage{pdflscape}
\usepackage{adjustbox}
\usepackage{xcolor}
\usepackage{threeparttable}

\usepackage{graphicx}	
\usepackage{amsmath}	
\usepackage{amssymb}	

\title[C-BASS Northern Sources]{The C-Band All-Sky Survey (C-BASS): Total intensity point-source detection over the northern sky}

\author[R.\,D.\,P. Grumitt et al.]{R.D.P. Grumitt,$\!^{1}$\thanks{E-mail: richard.grumitt@physics.ox.ac.uk}
Angela C. Taylor,$^{1}$
Luke Jew,$\!^{1}$
Michael E. Jones,$\!^{1}$
C. Dickinson,$\!^{2,4}$
\newauthor
A. Barr,$\!^{2}$
R.\,Cepeda-Arroita,$\!^{2}$
H.\,C. Chiang,$\!^{3}$
S.\,E. Harper,$\!^{2}$
H.\,M. Heilgendorff,$\!^{5}$
\newauthor
J.\,L. Jonas,$\!^{6,7}$
J.\,P. Leahy,$\!^{2}$
J. Leech,$\!^{1}$
T.\,J. Pearson,$\!^{4}$
M.\,W. Peel,$\!^{8,9}$
\newauthor
A.\,C.\,S. Readhead,$\!^{4}$
J. Sievers$\!^{3}$
\\
$^{1}$Sub-department of Astrophysics, University of Oxford, Denys Wilkinson Building, Keble Road, Oxford OX1 3RH, UK\\
$^{2}$Jodrell Bank Centre for Astrophysics, Alan Turing Building, Department of Physics and Astronomy, School of Natural Sciences,\\ 
The University of Manchester, Oxford Road, Manchester, M13 9PL, U.K. \\
$^{3}$Department of Physics, McGill University, 3600 Rue University, Montr\'{e}al, QC H3A 2T8, Canada \\
$^{4}$Cahill Centre for Astronomy and Astrophysics, California Institute of Technology, Pasadena, CA 91125, USA \\
$^{5}$Astrophysics \& Cosmology Research Unit, School of Mathematics, Statistics \& Computer Science, University of KwaZulu-Natal, \\Westville Campus, Private Bag X54001, Durban 4000, South Africa\\
$^{6}$Department of Physics and Electronics, Rhodes University, Grahamstown, 6139, South Africa \\
$^{7}$South African Radio Astronomy Observatory, 2 Fir Road, Observatory, Cape Town, 7925, South Africa \\
$^{8}$Instituto de Astrof\'{i}sica de Canarias, E-38205 La Laguna, Tenerife, Spain\\
$^{9}$Departamento de Astrof\'{i}sica, Universidad de La Laguna (ULL), E-38206 La Laguna, Tenerife, Spain\\
}

\date{Accepted XXX. Received YYY; in original form ZZZ}

\pubyear{2019}

\begin{document}
\label{firstpage}
\pagerange{\pageref{firstpage}--\pageref{lastpage}}
\maketitle

\begin{abstract}
We present a point-source detection algorithm that employs the second order Spherical Mexican Hat wavelet filter (SMHW2), and use it on C-BASS northern intensity data to produce a catalogue of point-sources. This catalogue allows us to cross-check the C-BASS flux-density scale against existing source surveys, and provides the basis for a source mask which will be used in subsequent C-BASS and cosmic microwave background (CMB) analyses. The SMHW2 allows us to filter the entire sky at once, avoiding complications from edge effects arising when filtering small sky patches. The algorithm is validated against a set of Monte Carlo simulations, consisting of diffuse emission, instrumental noise, and various point-source populations. The simulated source populations are successfully recovered. The SMHW2 detection algorithm is used to produce a $4.76\,\mathrm{GHz}$ northern sky source catalogue in total intensity, containing 1784 sources and covering declinations $\delta\geq-10^{\circ}$. The C-BASS catalogue is matched with the Green Bank 6\,cm (GB6) and Parkes-MIT-NRAO (PMN) catalogues over their areas of common sky coverage. From this we estimate the $90$ per cent completeness level to be approximately $610\,\mathrm{mJy}$, with a corresponding reliability of $98$ per cent, when masking the brightest $30$ per cent of the diffuse emission in the C-BASS northern sky map. We find the C-BASS and GB6 flux-density scales to be consistent with one another to within approximately $4$ per cent.
\end{abstract}




\section{Introduction}\label{sec: intro}   

C-BASS is an experiment to observe the whole sky in total intensity and polarization at $4.76\,\mathrm{GHz}$, at $45\,\mathrm{arcmin}$ angular resolution \citep{2018MNRAS.480.3224J}. The primary purpose of the experiment is to provide data on synchrotron emission for CMB polarization experiments at low frequencies compared to the peak of the CMB, whilst avoiding significant de-polarization due to Faraday rotation. C-BASS will also provide improved measurements of low-frequency emission components, enabling a detailed study of the Galactic Magnetic Field. The northern sky survey has now been completed, with detailed analysis of these data currently underway. In this paper we present a $4.76\,\mathrm{GHz}$ northern sky point-source catalogue, produced using a point-source detection algorithm employing the SMHW2.

Extragalactic radio sources are a key contaminant in CMB studies, with their detection and removal being an important step in CMB component separation \citep{1994ApJ...424....1B, 2001MNRAS.327L...1T, 2003MNRAS.342..915W, 2007MNRAS.379.1442W, deZotti2010}. For future CMB experiments, it has been shown that point-source contamination has the potential to significantly complicate measurements of the $B$-mode polarization power spectrum on scales $\ell\gtrsim 50$ \citep{2002A&A...396..463M, 2013MNRAS.432..728C, 2018ApJ...858...85P, 2018JCAP...04..023R}. Given the large beam of C-BASS, the instrument is not well suited to point-source detection, with detections of all but the brightest sources ($\gtrsim 1\,\mathrm{Jy}$) being limited by source confusion and diffuse emission. For the C-BASS analysis, performing our own dedicated source detection allows us to carry out important data quality checks on the C-BASS flux-density scale and pointing accuracy, by comparing the C-BASS catalogue with pre-existing catalogues around $4.76\,\mathrm{GHz}$. The resulting catalogue will also allow us to construct an accurate mask of bright source emission in the C-BASS maps, which will be key to avoid biasing CMB component separation analyses with C-BASS. Fainter sources can either be subtracted using deeper, pre-existing source catalogues, or can be treated statistically at the power spectrum level. Assuming sources down to some flux density $S_{\mathrm{max}}$ have been removed, and that sources are Poisson-distributed on the sky, the point-source contribution to the power spectrum is given by
\begin{equation}
C_{\ell}^{\mathrm{PS}} = \int_{0}^{S_{\mathrm{max}}}\frac{\mathrm{d}N}{\mathrm{d}S}S^{2}\,\mathrm{d}S,
\label{PS power spectrum}
\end{equation}
where $S$ is the flux density and $\mathrm{d}N/\mathrm{d}S$ is the differential source count, i.e., the number of sources per unit flux density, per unit steradian \citep{1996MNRAS.281.1297T}.

Numerous source-detection algorithms have been developed for CMB experiments, see e.g., \cite{1998ApJ...500L..83T, 2003MNRAS.344...89V, 2009MNRAS.395..649A, 2013ApJS..208...20B, 2009MNRAS.394..510H, 2012MNRAS.427.1384C, 2006MNRAS.370.2047L, 2006MNRAS.369.1603G, 2014A&A...571A..28P, 2016A&A...594A..26P}. The Planck Catalogue of compact sources (PCCS) used an algorithm based on filtering small patches of sky using the flat-sky second-order Mexican Hat wavelet as an approximation to a matched filter, followed by threshold detection based on the signal-to-noise ratio (SNR). Here we employ a similar scheme but using the equivalent spherical function on the whole sky at once, implemented in the Healpix pixelization scheme \citep{2005ApJ...622..759G}. Previous applications of the first-order SMHW (SMHW1) to searches for non-gaussianity can be found in \cite{2001MNRAS.326.1243C, 2002MNRAS.336...22M, 2011MNRAS.412.1038C}, and to point-source detection in \cite{2003MNRAS.344...89V}. The details of this implementation are discussed in Section \ref{sec: cbass algorithm}.

The outline of this paper is as follows: In Section \ref{sec: pre-existing cats} we summarize pre-existing source catalogues directly relevant to our C-BASS analysis. In Section \ref{sec: cbass algorithm} we discuss the C-BASS point-source detection algorithm, giving an overview of the method and the Monte Carlo simulations used to validate the algorithm. In Section \ref{sec: cbass catalogue} we discuss the C-BASS total intensity, northern sky point-source catalogue obtained using our detection algorithm. We compare the C-BASS catalogue with the GB6 and PMN catalogues, and calculate the differential source counts for the C-BASS sources as a cross-check on the statistical properties of the bright source population. We summarize our results in Section \ref{sec: conclusions}. Whilst analysis of polarized sources is important for future CMB polarization studies, this falls beyond the scope of this paper where we focus on the C-BASS total intensity results. In addition, due to the low level of source polarization, only a small number of polarized sources ($\mathcal{O}(10)$) will be detected.
\section{Pre-Existing Source Catalogues}\label{sec: pre-existing cats}

In order to construct an accurate template for masking out the point sources in the C-BASS maps it is necessary to construct a point-source catalogue based on the C-BASS observations themselves and any useful information that can be gleaned from other more sensitive, higher resolution surveys. Relevant to our work in this paper are the GB6 \citep{1996ApJS..103..427G}, PMN \citep{1993AJ....105.1666G, 1994ApJS...91..111W, 1994ApJS...90..179G, 1995ApJS...97..347G, 1996ApJS..103..145W}, Effelsberg S5 \citep{1981AJ.....86..854K}, RATAN-600 \citep{2007ARep...51..343M} and Combined Radio All-Sky Targeted Eight-GHz Survey (CRATES) \citep{2007ApJS..171...61H} catalogues. These catalogues are primarily used in this paper to make comparisons with the C-BASS catalogue, for the purposes of data validation and estimation of the statistical properties of the C-BASS catalogue. However, the catalogues are also useful for characterizing the faint point-source population in any C-BASS analysis (i.e., sources with flux-densities below $\sim 1\,\mathrm{Jy}$), where reliable extraction from the C-BASS map becomes more challenging. These radio surveys are summarized in Table \ref{tab: radio surveys}. 

\begin{table*}
    \centering
    \caption{Summary of radio surveys relevant to the work in this paper. In stating the sky coverage, $\delta$ denotes declination, and $b$ denotes Galactic latitude.}
    \label{tab: radio surveys}
    \begin{threeparttable}
    \begin{tabular}{lcccccc}
        \hline
         Survey Name & Reference & Frequency & Sky & FWHM & Flux Limit & Number of \\
         & & [GHz] & Coverage & [arcmin] & [mJy] & Sources\\
         \hline
         C-BASS & \cite{2018MNRAS.480.3224J} & 4.76 & $-90^{\circ}\leq\delta\leq 90^{\circ}$$^{a}$ & $45$ & $\sim 500^{b}$ & 1784\\
         GB6 & \cite{1996ApJS..103..427G} & 4.85 & $0^{\circ}\leq\delta\leq 75^{\circ}$ & 3 & $\sim 18$ & 75,162\\
         PMN & \cite{1996yCat.8038....0W} & 4.85 & $-87.5^{\circ}\leq\delta\leq 10^{\circ}$ & 5 & $\sim 35$ & 50,814\\
         Effelsberg S5 & \cite{1981AJ.....86..854K} & 4.9 & $70^{\circ}\leq\delta\leq 90^{\circ}$ & 2.7 & $\sim 250$ & 476\\
         RATAN-600 & \cite{2007ARep...51..343M} & $4.8\: (1.1 - 21.7)^{c}$ & $75^{\circ}\leq\delta\leq 88^{\circ}$ & $0.67\times 6.6^{d}$ & $(200)^{e}$ & 504\\
         CRATES & \cite{2007ApJS..171...61H} & $8.4\: (4.85)^{f}$ & $\left|b\right|>10^{\circ}$ & 2.4 & $\sim 65$ & 11,131\\
         \hline
    \end{tabular}
    \begin{tablenotes}
    \item $^{a}$ Here we state the sky coverage of the whole C-BASS experiment. The results concerning point sources in this paper were obtained for the C-BASS northern intensity data, covering declinations $-10^{\circ}\leq\delta\leq 90^{\circ}$. 
    \item $^{b}$ For C-BASS we state the value $3.5\langle\sigma\rangle$, where $\langle\sigma\rangle$ is the mean background fluctuation level found across the map, applying the CG30 mask described in Section \ref{subsec: val comp and positions}. Details on the estimation of background fluctuation levels are given in Section \ref{subsec: algorithm structure}.
    \item $^{c}$ The RATAN-600 catalogue covers 6 frequencies from $1.1 - 21.7\,\mathrm{GHz}$, including $4.8\,\mathrm{GHz}$.
    \item $^{d}$ We state $\mathrm{FHWM_{RA}}\times\mathrm{FWHM_{\delta}}$ for RATAN-600, determined from the values given in \cite{1999A&AS..139..545K}.
    \item $^{e}$ The RATAN-600 catalogue was produced by pre-selecting NVSS sources with $S_{1.4\,\mathrm{GHz}}>200\,\mathrm{mJy}$.
    \item $^{f}$ The CRATES catalogue is primarily at $8.4\,\mathrm{GHz}$, with $4.85\,\mathrm{GHz}$ sources used as the basis for observations of $8.4\,\mathrm{GHz}$ counterparts. The properties of these $4.85\,\mathrm{GHz}$ sources are provided with the CRATES catalogue. Further observations were also made at $4.85\,\mathrm{GHz}$ to fill in gaps at $\delta>88^{\circ}$.
    \end{tablenotes}
    \end{threeparttable}
\end{table*}

The GB6 and PMN $4.85\,\mathrm{GHz}$ source catalogues cover declinations $-87.5^{\circ}\leq\delta\leq 75^{\circ}$. The GB6 catalogue was produced using the NRAO seven-beam receiver on the $91\,\mathrm{m}$ telescope, and the PMN catalogue was produced using the Parkes $64\,\mathrm{m}$ radio telescope. The GB6 catalogue has a flux-density limit of approximately $18\,\mathrm{mJy}$, whilst the PMN catalogue has an average flux-density limit of approximately $35\,\mathrm{mJy}$ over the sky. The GB6 and PMN catalogues provide far deeper flux-density coverage than can be achieved with C-BASS, given the differing resolutions and hence confusion levels. However, it is still necessary for us to obtain our own source catalogue, such that we can construct an accurate mask for the brightest sources in the C-BASS maps, and account for source variability between the C-BASS and GB6/PMN surveys. The GB6 and PMN catalogues remain useful in accounting for fainter sources (below $\sim 1\,\mathrm{Jy}$) in any C-BASS analysis.

Source catalogues covering the North Celestial Pole (NCP) region are more limited, with GB6 only covering declinations up to $\delta = 75^{\circ}$. The Effelsberg S5 catalogue covers declinations $\delta\geq 70^{\circ}$ and is complete down to $250\,\mathrm{mJy}$ \citep{1981AJ.....86..854K}. In comparing to the S5 catalogue there are significant issues from the variability of flat-spectrum sources, given the large separation in time between the S5 survey and the C-BASS survey. The RATAN-600 catalogue includes measurements at $4.8\,\mathrm{GHz}$, and observed 504 sources in the NCP region with NRAO VLA Sky Survey (NVSS) flux-densities, $S_{1.4\,\mathrm{GHz}}\,\geq 200\mathrm{mJy}$ \citep{2001A&A...370...78M, 2007ARep...51..343M}. The RATAN-600 catalogue was used in a previous analysis of diffuse emission in the NCP region with C-BASS in \cite{2019MNRAS.485.2844D}. Whilst this catalogue provides deeper flux-density coverage than C-BASS, it was produced by pre-selecting sources for study from the NVSS catalogue at $1.4\,\mathrm{GHz}$ \citep{1998AJ....115.1693C}. This can potentially miss rising-spectrum sources that would otherwise be observable in the C-BASS catalogue. Given the C-BASS flux-density limit of approximately $500\,\mathrm{mJy}$, we can use the NVSS source counts and the $1.4 - 4.85\,\mathrm{GHz}$ spectral index distributions in \cite{2011A&A...533A..57T} to estimate that there may be $\mathcal{O}(1)$ sources that could be observed by C-BASS, whilst being missed by RATAN-600.

It is also worth noting the CRATES catalogue \citep{2007ApJS..171...61H}. CRATES is an $8.4\,\mathrm{GHz}$ catalogue of flat-spectrum sources with flux-densities $S_{\nu=4.85\,\mathrm{GHz}}>65\,\mathrm{mJy}$, covering Galactic latitudes $|b|>10^{\circ}$. The catalogue therefore serves as a useful proxy for flat-spectrum sources in the C-BASS catalogue, which are the primary contributor to source variability. \cite{2009AJ....138.1032H} made additional observations of the NCP region at declinations, $\delta>88^{\circ}$ to supplement the original CRATES catalogue. The purpose of this was to bring the flux-density limit in this region down to the CRATES flux-density limit of $\sim 65\,\mathrm{mJy}$. Three sources were observed in this region at $4.85\,\mathrm{GHz}$, with flux-densities of $67\,\mathrm{mJy}$, $58\,\mathrm{mJy}$ and $142\,\mathrm{mJy}$.  
\section{C-BASS Source Detection Algorithm}\label{sec: cbass algorithm}

To detect sources in a sky map we need to remove obscuring diffuse emission and noise. For a source with a known point-spread function (PSF) embedded in additive noise, the matched filter (MF) is the optimal filter that can be applied to maximize the source SNR. The matched filter is given by
\begin{equation}
    \Psi_{\mathrm{MF}}(k) = \left[2\pi\int\mathrm{d}k\,k\frac{\tau^{2}(k)}{P(k)}\right]^{-1}\frac{\tau(k)}{P(k)},
\end{equation}
where $k$ is the Fourier wavenumber, $\tau(k)$ is the source profile and $P(k)$ is the power spectrum of the unfiltered map \citep{1998ApJ...500L..83T, 2006MNRAS.370.2047L}. However, the calculation of the MF involves a number of complications. Chiefly, we are required to make a noisy estimate of the power spectrum from our unfiltered map, and integrate it. In constructing the PCCS, it was found that the Mexican Hat wavelet of the second kind (MHW2) achieved similar performance to the MF \citep{2006MNRAS.370.2047L, 2014A&A...571A..28P, 2016A&A...594A..26P}. For the present analysis we adapt the \textit{Planck} algorithm, using instead the SMHW2 in place of the flat-space MHW2. The SMHW2 is straightforward to calculate, enables us to filter the entire sky at once, and allows us to optimize a few free parameters as opposed to a noisy estimate of the full noise power spectrum.

\subsection{Source Detection Algorithm}\label{subsec: algorithm structure}

\begin{figure*}
\centering
\includegraphics[width=\textwidth]{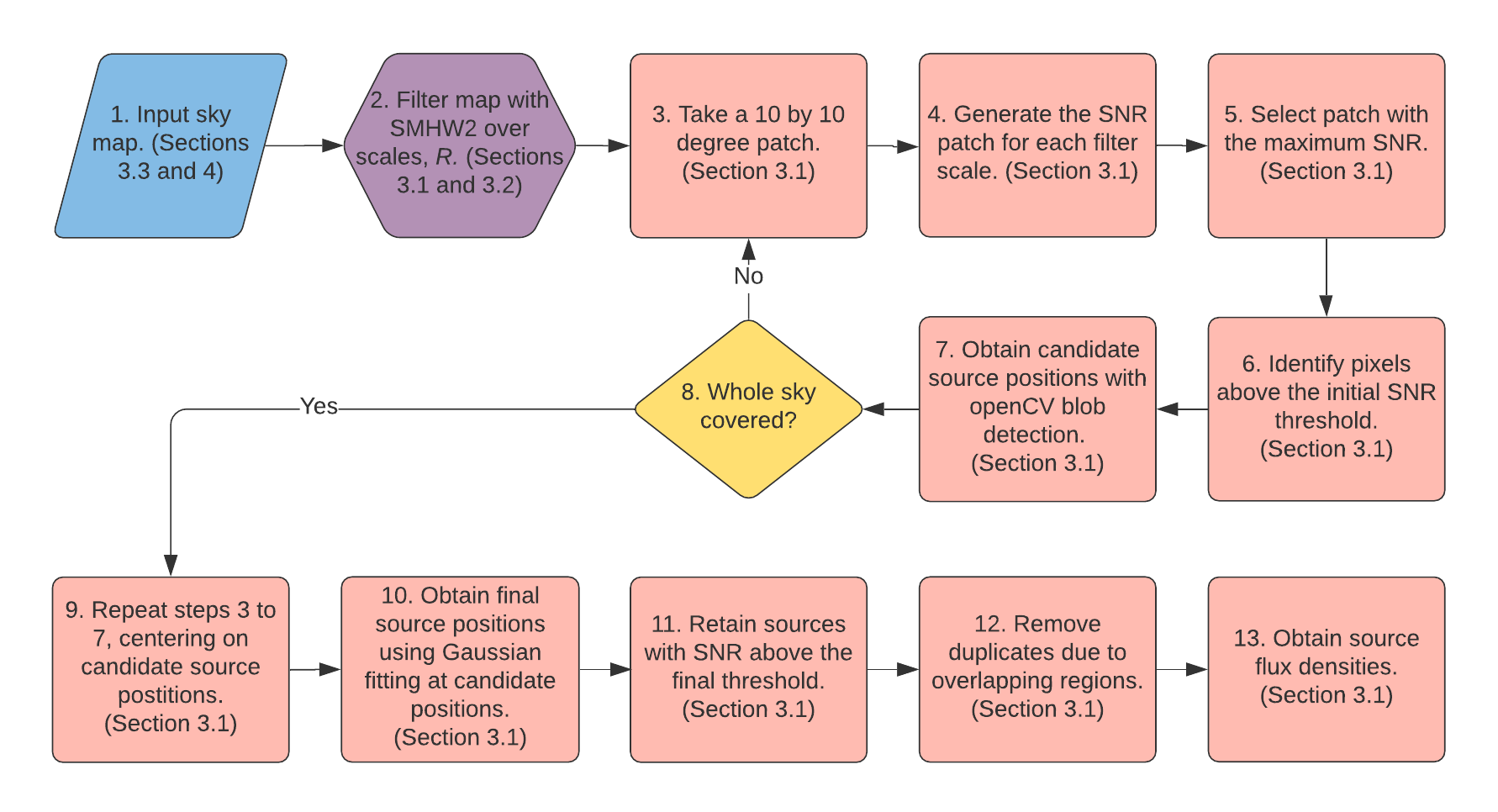}
\caption{Flowchart for the C-BASS point-source detection algorithm. The SMHW2 filter is applied over the entire sky, whilst identification of point sources is performed on $10^{\circ}\times 10^{\circ}$ sky regions.}
\label{fig:algo flow}
\end{figure*}

Given a sky map consisting of point sources, diffuse emission and instrumental noise, we can enhance the SNR of point sources in the map by filtering with the SMHW2. We perform this filtering at a range of filter scales $R$, to maximize the number of sources we extract from the sky maps. In changing $R$, we effectively change the extent to which we down-weight large-scale $\ell$-modes that are dominated by diffuse emission, and also small scales dominated by instrumental noise. This is particularly important in regions close to the Galactic plane where diffuse emission is very strong, meaning we must down-weight large scale modes harshly in order to extract point sources. Conversely, in regions with little diffuse emission we may wish to be less extreme in our down-weighting, so that we do not excessively reduce point-source power and subsequently miss detection of fainter sources in these regions. We discuss the form of the SMHW2 filter and the effect of the filter scale, $R$ in detail in Section \ref{subsec: smhw2}.

The source-detection algorithm is described below, with the algorithm flowchart displayed in Fig. \ref{fig:algo flow}. 

\begin{figure*}
\centering
\subfigure[Input sky]{\includegraphics[width=\textwidth]{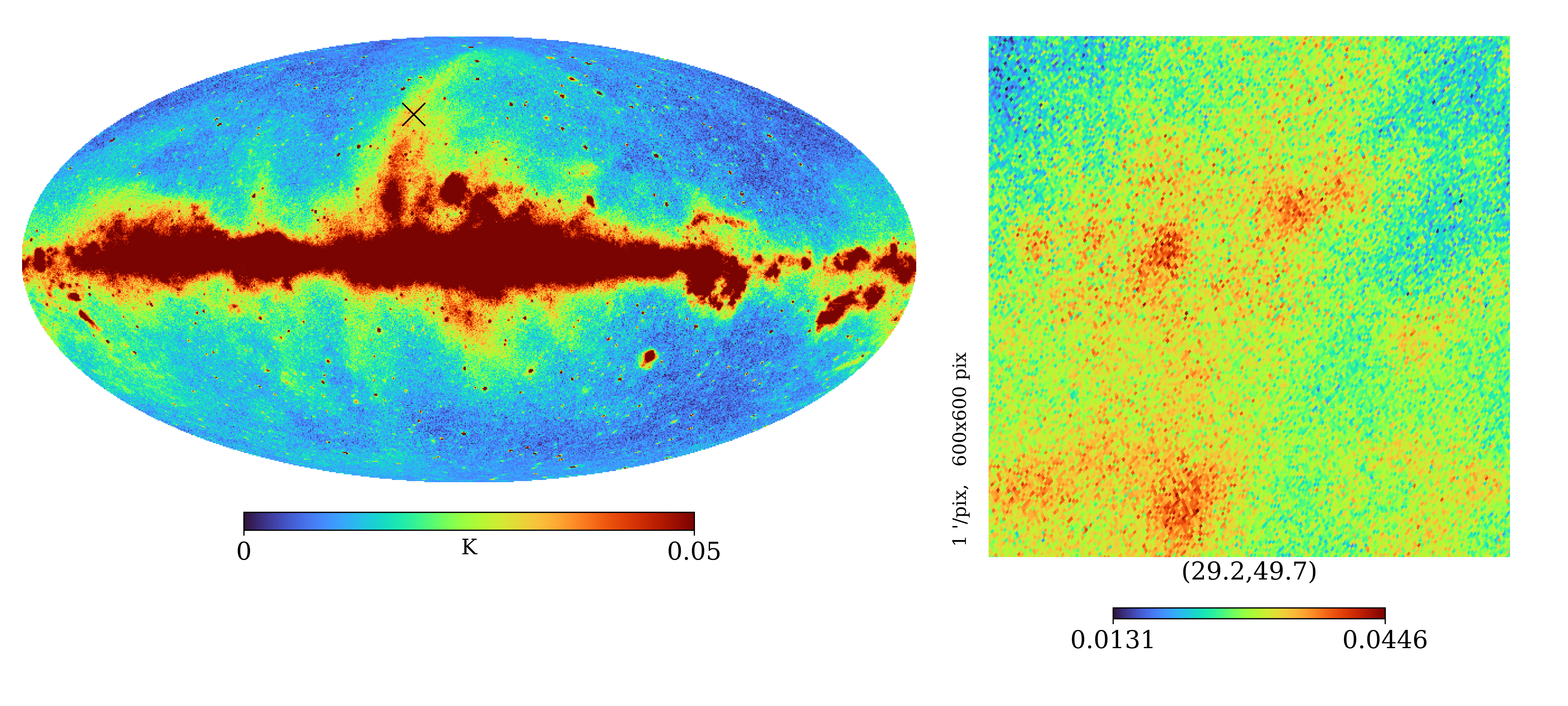}}
\subfigure[Filtered sky]{\includegraphics[width=\textwidth]{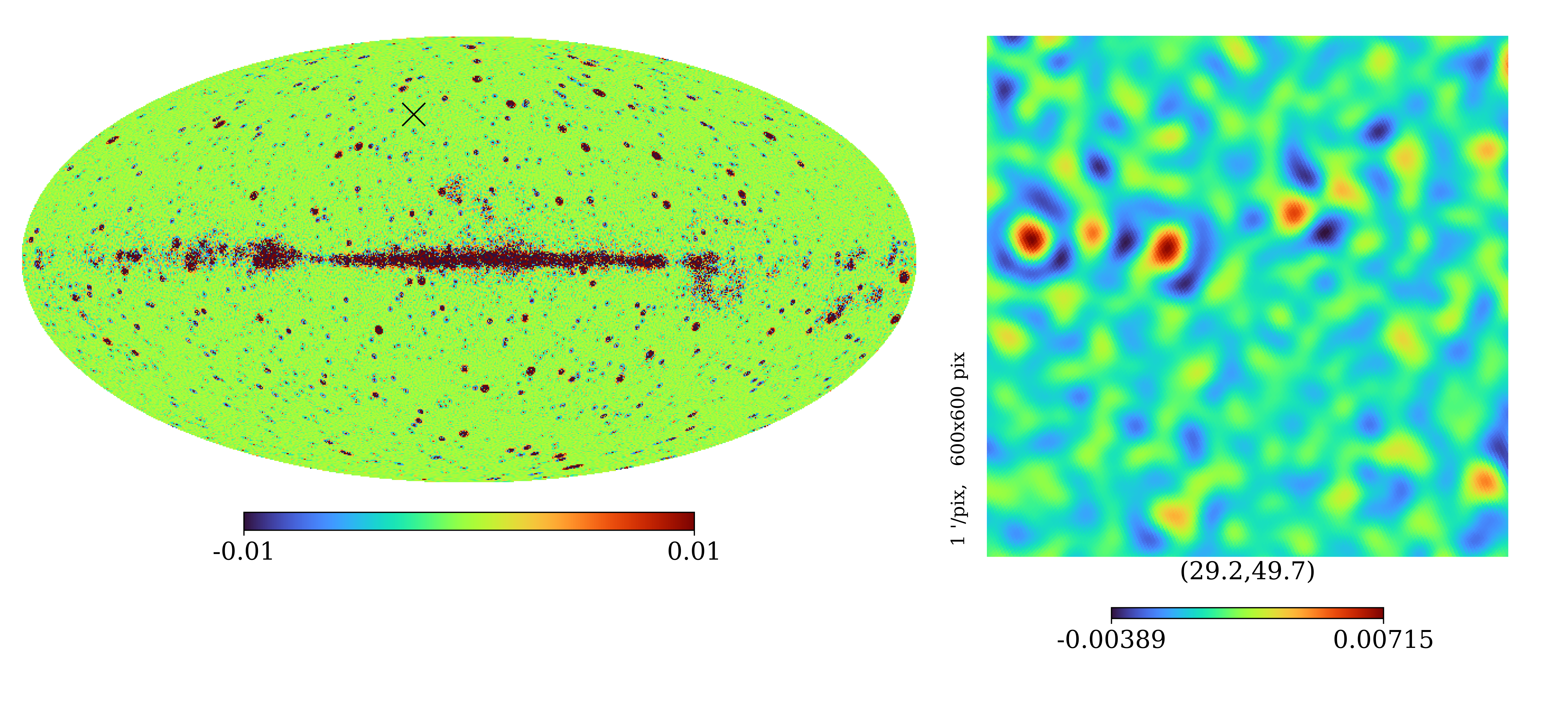}}
\caption{Panel (a): An input $4.76\,\mathrm{GHz}$ simulated sky consisting of diffuse emission generated using \textsc{PySM}, instrumental noise generated from a white-noise realization of the C-BASS sensitivity map, along with a point-source population generated by scattering the GB6, PMN and RATAN-600 sources at random positions over the sky. Alongside this we display a zoom-in centered on the location of the black cross on the map. Panel (b): The same map after being convolved with the SMHW2 filter, removing the large scale diffuse emission and leaving behind a sky filled with point-sources. As before, we display a zoom-in of the filtered map alongside it, centred on the location of the black cross. It is important to note here that not all the diffuse emission has been removed at low Galactic latitudes. In regions of strong diffuse emission the source detection algorithm is significantly less reliable. In panel (b) it is apparent that the SMHW2 filtering can still leave a significant sidelobe around bright sources. This can lead to issues with the spurious tagging of sidelobe peaks as sources, particularly for the brightest sources on the sky (i.e., those with flux-densities $\gtrsim 20\,\mathrm{Jy}$). This problem can largely be side-stepped by allowing for larger exclusion zones around the brightest sources, as was the case with catalogues such as GB6, removing apparent detections in the first sidelobe of such sources.}
\label{fig:sky sims}
\end{figure*}

\begin{figure*}
\centering
\subfigure[SNR patch]{\includegraphics[width=0.32\textwidth]{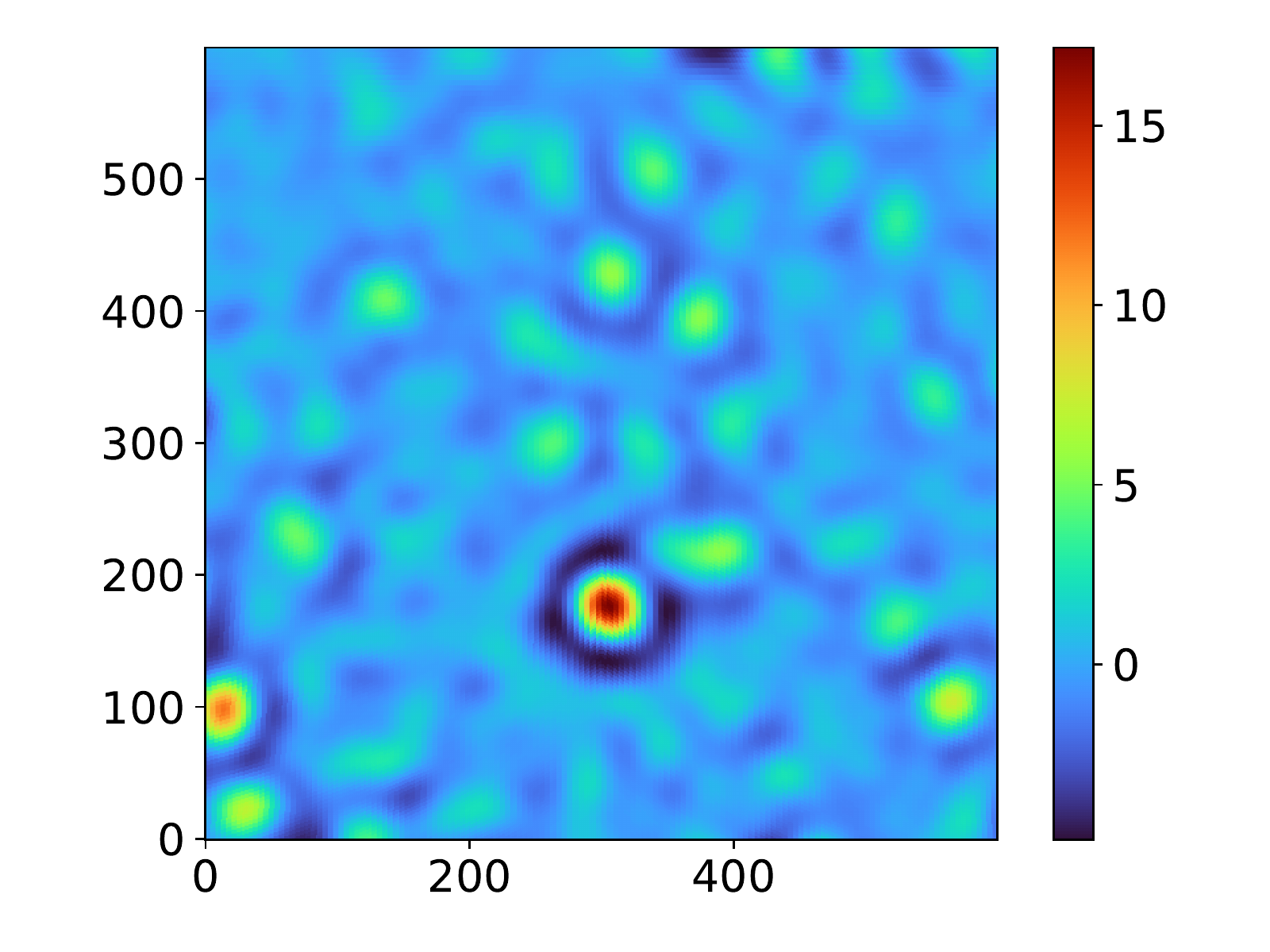}}
\subfigure[Tagged patch]{\includegraphics[width=0.32\textwidth]{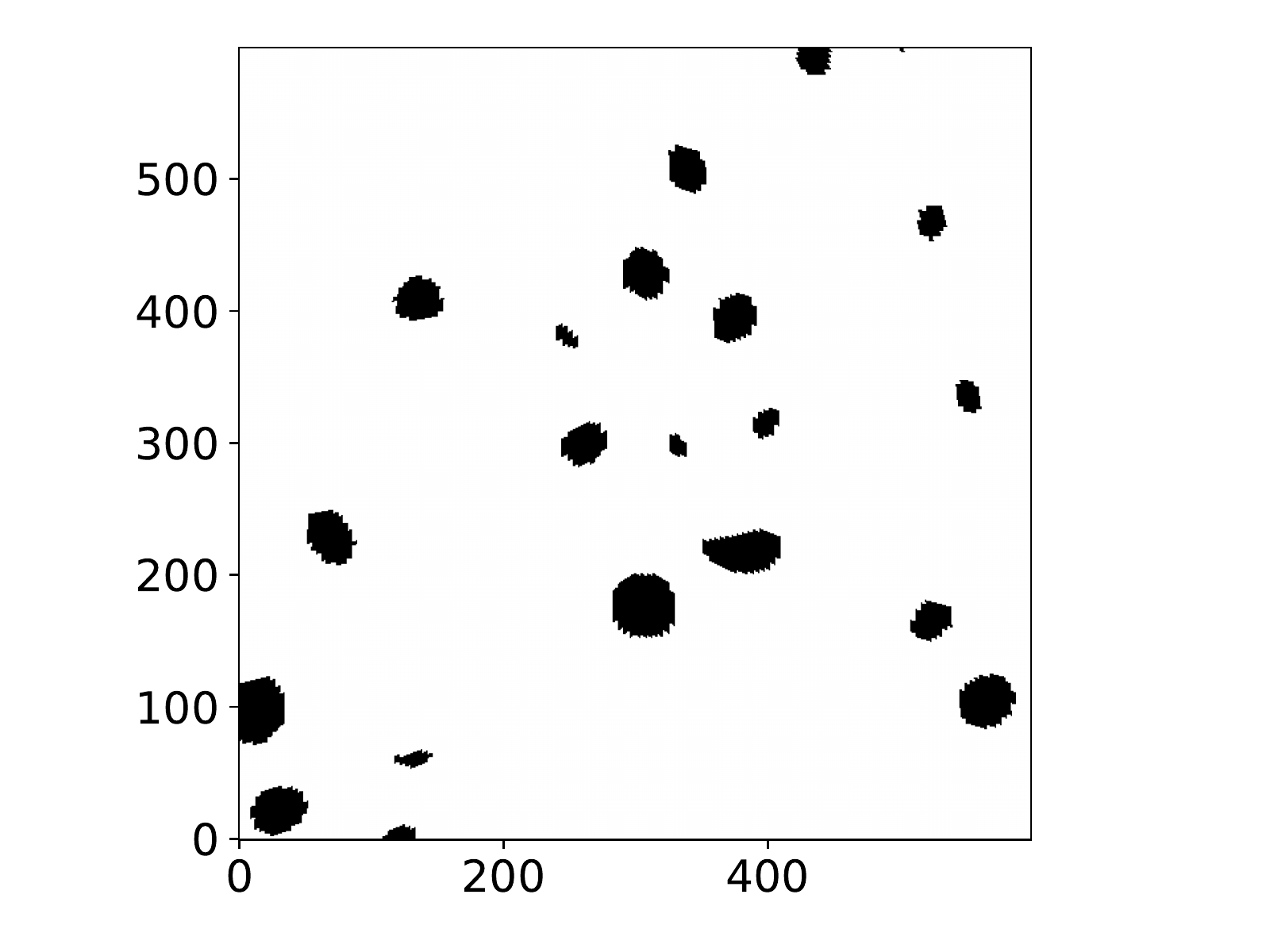}}
\subfigure[Detected source locations]{\includegraphics[width=0.32\textwidth]{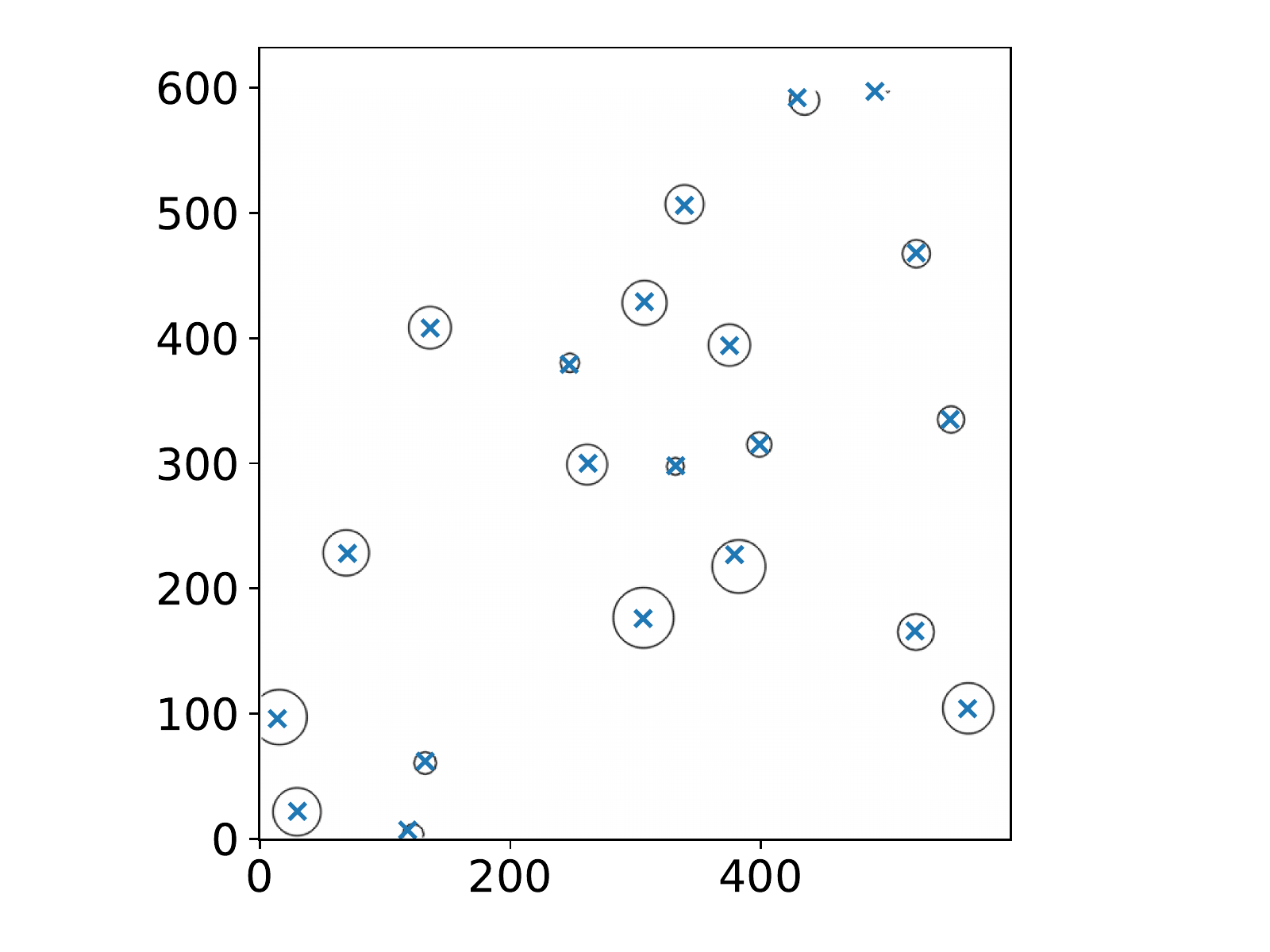}}
\caption{Images illustrating the pixel tagging process. Panel (a): An example $10^{\circ}\times 10^{\circ}$ SNR patch, produced by dividing the filtered sky patch by the MAD value of the pixels in the patch. Panel (b): The corresponding patch of tagged and un-tagged pixels. The tagged pixels are shown in \textit{black} and un-tagged pixels in \textit{white}, obtained by applying a threshold of $\mathrm{SNR}\geq 3.5$ to the SNR patch. Panel (c): An image showing the detected source locations for this patch. We use the \textsc{OpenCV} blob detection algorithm to obtain preliminary source locations from this patch. These blob detections are shown as black circles centred on the preliminary source locations. These position estimates are then refined by fitting an elliptical Gaussian at the preliminary source locations, with the source positions obtained from the Gaussian fit shown as blue crosses.}
\label{fig:algo zoom}
\end{figure*}

\begin{enumerate}
	\item We begin by taking the spherical harmonic transform of our sky map, obtaining the $a_{\ell m}$ coefficients. We filter the map by weighting the $a_{\ell m}$ coefficients with the SMHW2 window function, $w_{\ell}^{S^2}\left(R\right)$ at some user-defined set of filter scales, i.e. we calculate
		\begin{equation}
			f_{\ell m}\left(R\right)=a_{\ell m}w_{\ell}^{S^2}\left(R\right).
		\label{filtered alm}
		\end{equation}
    The effect of this filtering is demonstrated in Fig. \ref{fig:sky sims}. Here we display a simulated sky consisting of diffuse emission, instrumental noise and point-sources, along with the corresponding filtered sky with scale $R=1$. The details of the production of this simulated sky are discussed in Section \ref{validation}. The filter has removed diffuse emission and enhanced the SNR of point-sources across the sky.
	\item We break the maps, filtered at a range of $R$ scales, into overlapping $10^{\circ}\times 10^{\circ}$ patches on the sky and calculate the corresponding SNR patches. We used $R$ scales in the range $0.5\leq R\leq2$ in steps of 0.1. We choose $10^{\circ}\times 10^{\circ}$ patches such that the flat-sky approximation holds, whilst capturing a sufficient sample of the sky background in the given sky region. We set the overlap to $5^{\circ}$ in each direction to ensure that we obtain accurate positions for sources detected close to patch edges. The typical background fluctuation level in each patch is estimated by calculating the median absolute deviation (MAD) of the pixels, defined as
	\begin{equation}
	    \mathrm{MAD}(\mathbfit{X}) = \mathrm{median}\left(\left|X_{i} - \mathrm{median}(\mathbfit{X})\right|\right),
	\end{equation}
	where $\mathbfit{X}$ is the set of pixel values. The MAD estimator avoids sensitivity to outlier pixel values, which can be caused by the presence of point sources in the sky patch. For each sky patch, we then select the filter scale that gives the maximal SNR for the patch i.e., the maximum of the peak pixel value divided by the MAD estimate.
    \item Having selected the value of $R$ that maximizes the patch SNR, an initial SNR threshold is applied to identify potential point sources. This produces a patch of tagged and un-tagged pixels. The tagging is performed on these small, overlapping regions so as to account for the varying background properties across the sky i.e., variations in residual diffuse emission and noise.
	\item Given our patch of tagged pixels, we then use the \textsc{OpenCV} Simple Blob Detector\footnote{\url{https://opencv.org/}} to obtain candidate source locations. The Simple Blob Detector algorithm works by applying thresholds to an input image and determining the centres of connected pixels, or sources. In addition to providing candidate source locations, this software allows for the filtering of detected sources according to their geometric properties i.e., their size, circularity, convexity and inertia ratio. For our purposes, we set a maximum size limit to help prevent spurious tagging of bright diffuse emission and disallow detections within one beam FWHM of one another. For a detailed discussion of \textsc{OpenCV} and the Simple Blob Detector algorithm we refer the reader to \cite{kaehler2016learning}. The processing steps involved in studying these small sky regions are illustrated in Fig. \ref{fig:algo zoom}, where we display the typical output from the steps outlined above.
	\item Given a set of tentative source locations obtained over the whole sky, we now repeat steps (ii) and (iii), this time centering on the tentative source locations. In order to refine the source position estimates, we fit an elliptical Gaussian to a $20\,\mathrm{arcmin}$ square region of the wavelet-filtered map, centered on the candidate location obtained using \textsc{OpenCV}. Whilst the C-BASS beam and SMHW2 filter are not Gaussian, by fitting to this small region around the candidate location the Gaussian fit is sufficient to make the necessary refinements ($\sim 1\,\mathrm{arcmin}$) to our estimates of source peak locations.
	\item A final SNR thresholding is applied by dividing the peak value of the source in the wavelet filtered map by the MAD value of pixels in an annulus centered on the source, with an inner radius of $3^{\circ}$ and outer radius of $5^{\circ}$. The annulus was chosen to avoid the first sidelobe in the C-BASS beam, whilst capturing the local background fluctuations around the source. Sources are retained if this new SNR estimate exceeds some final threshold. The SMHW2 is constructed so as to preserve point-source amplitudes after being convolved with the C-BASS beam. As with the PCCS, we convert these wavelet amplitudes to $\mathrm{Jy}$ and report them as auxiliary flux-density (DETFLUX) estimates alongside the primary estimates from aperture photometry described in step (vii). The DETFLUX estimates are discussed in more detail in Appendix \ref{appndx: catalogue sample}.
	\item The detected source locations from all of the filtered maps are combined and duplicates are removed from the catalogue, defined here to be any reported source positions that are within the beam FWHM of one another. For matched sources in our internal catalogue, we retain the source with the greatest detected SNR.
	\item Given the final set of source positions, source flux densities are obtained. As its primary method, the detection pipeline obtains flux densities using aperture photometry, adapting the method from \cite{2014A&A...571A..28P, 2015MNRAS.452.4169G}. We begin by converting the map to $\mathrm{Jy}\,\mathrm{pix}^{-1}$, and define an aperture of radius $45\,\mathrm{arcmin}$ around the source position, and an annulus of inner radius $3^{\circ}$ and outer radius $5^{\circ}$. The observed flux density is then given by
	\begin{equation}
	    S_{\mathrm{obs}} = \kappa\left(S_{\mathrm{ap}} - \bar{S}_{\mathrm{ann}}N_{\mathrm{ap}}\right),
	\end{equation}
	where $\kappa\approx 1.34$ is a correction applied to account for flux density missing from the aperture (calculated using Equation A.2 of \cite{2014A&A...571A..28P}), $S_{\mathrm{ap}}$ is the total flux density in the aperture, $\bar{S}_{\mathrm{ann}}$ is the median flux density in the annulus and $N_{\mathrm{ap}}$ is the number of pixels in the aperture. The uncertainty was estimated as
	\begin{equation}
	    \sigma(S_{\mathrm{obs}}) \approx \mathrm{MAD}(\mathbfit{X}_{\mathrm{ann}})\frac{N_{\mathrm{ap}}}{\sqrt{N_{\mathrm{ap}}'}},
	\end{equation}
	where $\mathrm{MAD}(\mathbfit{X}_{\mathrm{ann}})$ is the MAD value of pixels in the annulus and $N_{\mathrm{ap}}'$ is the number of beams inside the aperture. The scaling applied to $\mathrm{MAD}(\mathbfit{X}_{\mathrm{ann}})$ is to account for correlations in the background emission, approximately on the scale of the beam \citep{2015MNRAS.452.4169G}. 
\end{enumerate}

\subsection{The Spherical Mexican Hat Wavelet}\label{subsec: smhw2}

\begin{figure*}
\centering
\subfigure[Real space SMHW2]{\includegraphics[width=\columnwidth]{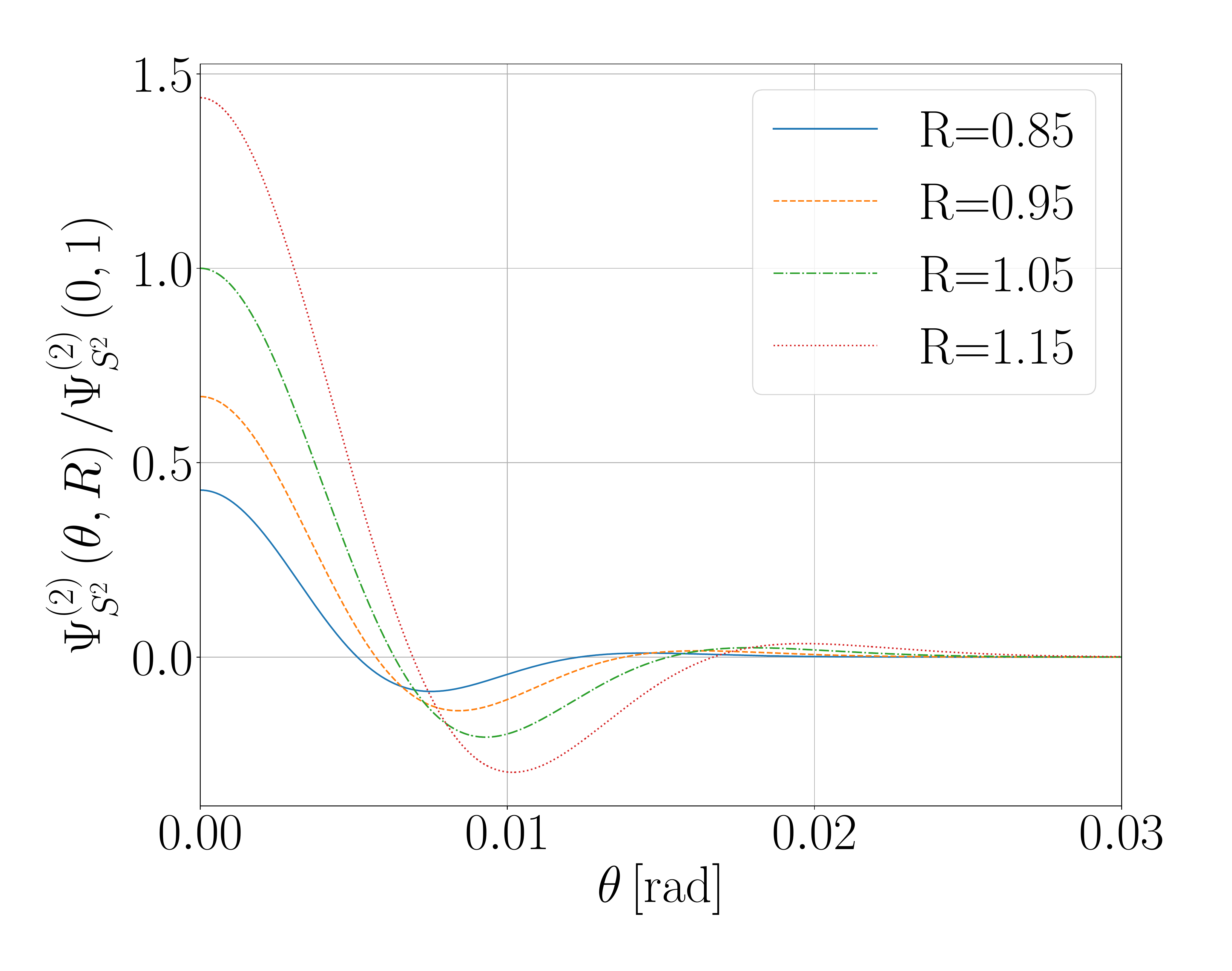}}
\subfigure[SMHW2 window function]{\includegraphics[width=\columnwidth]{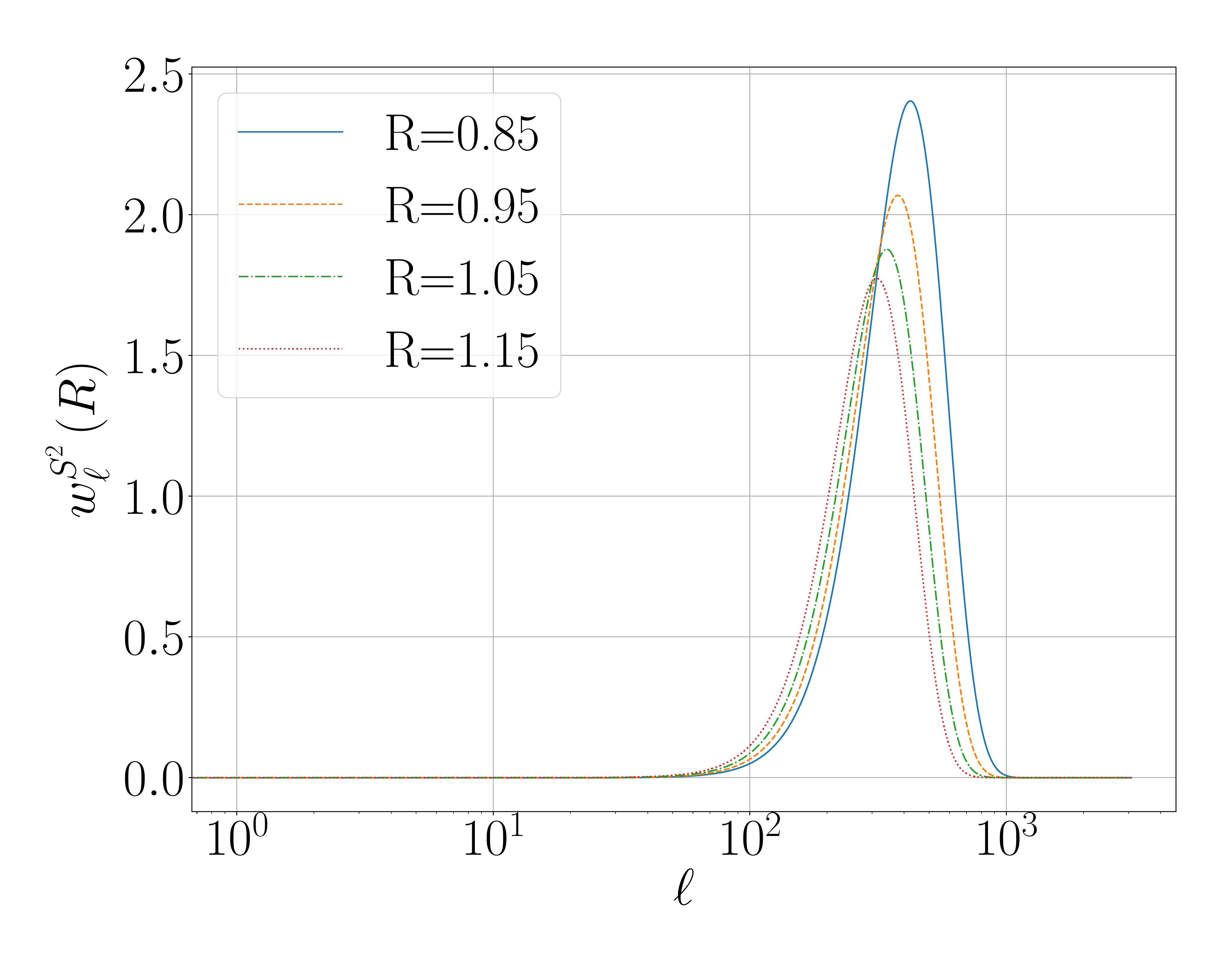}}
\caption{Panel (a): The real space SMHW2, shown for $\sigma$ corresponding to a Gaussian beam FWHM of $45\,\mathrm{arcmin}$ and $R\in\{0.85,0.95,1.05,1.15\}$. Panel (b): The corresponding window functions. The SMHW2 down-weights large and small scales, with changes in $R$ determining the severity with which we down-weight large- and small-scale emission. Whilst the C-BASS beam is not exactly Gaussian, by constructing the SMHW2 around these $\sim 45\,\mathrm{arcmin}$ scales we are able to filter the large- and small-scale modes around the characteristic beam scale, enhancing point-source SNR in the C-BASS sky map.}
\label{fig:1deg SMHW2}
\end{figure*}

The Mexican Hat Wavelet (MHW) of the $n^{\mathrm{th}}$ kind is defined in $\mathbb{R}^{2}$ (i.e., 2-dimensional Euclidean space) as
\begin{equation}
\Psi_{n}\left(\mathbfit{x}, R\right)\propto \Delta^{n}\left\{\exp\left(-\frac{\left|\mathbfit{x}\right|^{2}}{2\left(R\sigma\right)^{2}}\right)\right\},
\label{real MHWn}
\end{equation}
where $\mathbfit{x}$ is the position in the image plane, $\Delta$ is the usual Euclidean Laplacian operator, $\sigma$ is the Gaussian standard deviation and $R$ is the filter scale factor. In changing $R$, we change the characteristic scale of the filter and hence the extent to which the filter down-weights large- and small-scale modes. Applying this expression we can arrive at the real-space expression for the MHW2, given by
\begin{equation}
\Psi_{2}\left(\mathbfit{x}, R\right)\propto\left(8\left(R\sigma\right)^4-8\left(R\sigma\right)^{2}\left|\mathbfit{x}\right|^{2}+\left|\mathbfit{x}\right|^{4}\right)\exp\left(-\frac{\left|\mathbfit{x}\right|^{2}}{2\left(R\sigma\right)^{2}}\right).
\label{Real MHW2}
\end{equation}

In applying the MHW2 to producing the PCCS, the sky was divided up into small patches, taking flat projections and applying the MHW2 filter to them. SNR thresholding was applied to the filtered patches to extract the source positions and flux densities were then obtained for the detected sources. Here we employ the SMHW2, the equivalent of the MHW2 on $S^{2}$ (i.e., the 2-sphere). For a given filter scale, this allows us to filter the entire sky at once, avoiding complications from edge effects when filtering small patches. In dealing with the C-BASS northern sky map, we apply a $5^{\circ}$ $\cos^{2}$ apodization at the edge of our mask to mitigate edge effects. 

In \cite{Antoine1999} it was shown the continuous wavelet transform on $S^{2}$ can be constructed by taking the inverse stereographic projection of the $\mathbb{R}^{2}$ wavelet. This preserves the properties of the $\mathbb{R}^{2}$ wavelet and tends to the MHW in the small-angle limit. The spherical MHW of the $n^{\mathrm{th}}$ kind is given by
\begin{equation}
\Psi_{S^{2}}^{\left(n\right)}\propto\frac{1}{\cos^{4}\tfrac{\theta}{2}}\Psi_{n}\left(\left|\mathrm{x}\right|\equiv 2\tan\frac{\theta}{2}, R\right),
\label{SMHWn}
\end{equation}
where $\theta$ is the colatitude on the sphere \citep{2001MNRAS.326.1243C, 2002MNRAS.336...22M, 2003MNRAS.344...89V, 2011MNRAS.412.1038C}.

Substituting the MHW2 into Equation \ref{SMHWn}, we obtain the SMHW2 as follows,
\begin{multline}
\Psi_{S^{2}}^{\left(2\right)}\left(\theta, R\right)\propto\frac{1}{\cos^{4}\frac{\theta}{2}}\left[\left(R\sigma\right)^{4}-4\left(R\sigma\right)^{2}\tan^{2}\frac{\theta}{2}\right. \\ \left.
+2\tan^{4}\frac{\theta}{2}\right]\exp\left(-\frac{2\tan^{2}\frac{\theta}{2}}{\left(R\sigma\right)^{2}}\right).
\label{SMHW2}
\end{multline}
We normalize the SMHW2 such that we preserve point-source amplitude after convolving with the filter, i.e.
\begin{equation}
\int_{S^{2}}\mathcal{B}(\theta)\Psi_{S^{2}}^{\left(2\right)}\left(\theta, R\right)\sin\theta\,\mathrm{d}\theta\mathrm{d}\phi=1,
\label{SMHW2 norm}
\end{equation}
where $\mathcal{B}(\theta)$ is the C-BASS beam profile.

Given the analytic expression for the SMHW2 in real space on $S^{2}$, we can obtain the SMHW2 window function by calculating the Legendre transform of Equation \ref{SMHW2}, i.e. we calculate
\begin{equation}
w_{\ell}^{S^2}\left(R\right)=2\pi\int_{0}^{\pi}\Psi_{S^{2}}^{\left(2\right)}\left(\theta, R\right)P_{\ell}\left(\cos\theta\right)\sin\theta\,\mathrm{d}\theta,
\label{SMHW2 window}
\end{equation}
where $P_{\ell}\left(\cos\theta\right)$ is the Legendre polynomial of order $\ell$.

In Fig. \ref{fig:1deg SMHW2} we show the real-space SMHW2 and the corresponding window function, calculated for a $\sigma$ value corresponding to a beam FWHM of $45\,\mathrm{arcmin}$, with $R=\{0.85, 0.95, 1.05, 1.15\}$. The primary effect of the SMHW2 is to down-weight large scales, which are dominated by diffuse emission. In addition to this, the SMHW2 also acts to down-weight small-scale modes where instrumental noise becomes more significant. In the context of C-BASS, large-scale modes correspond to modes significantly larger than the C-BASS beam scale, $\ell\lesssim 100$, whilst small scales correspond to scales significantly smaller than the C-BASS beam scale, $\ell\gtrsim 1000$.

\subsection{Algorithm Validation}\label{validation}

We validate the algorithm by running it over 100 simulations of the sky at 4.76\,\textrm{GHz}. The simulated skies were generated at NSIDE=1024, where the number of equal-area pixels on the Healpix sphere is given by $12\times\mathrm{NSIDE}^{2}$. This high resolution was chosen to allow for the precise determination of source positions, with the C-BASS PSF over-sampled by a factor of $\sim 6.5$. The simulated sky includes diffuse emission generated by \textsc{PySM} \citep{2017MNRAS.469.2821T} and an instrumental noise realization. The noise maps were created by generating white noise realizations of the C-BASS sensitivity map (including a simulated sensitivity map for the southern sky survey). This allows us to include the declination dependent effects from the C-BASS scan strategy.

To the diffuse emission simulation we add a source population to the sky. These are created by taking source flux-densities from the PMN, GB6 and RATAN-600 surveys, summarised in Section \ref{sec: pre-existing cats}, and scattering the sources at random positions over the sky. Carrying this out 100 times we obtain 100 Monte Carlo simulations of the sky, consisting of the same diffuse background but with their source populations randomly distributed on the sky, and differing noise realizations. The simulated skies are all convolved with the C-BASS beam. 

The diffuse components included in the simulation are as follows:

\begin{enumerate}
	\item \textbf{Synchrotron Radiation}  This was generated using the de-sourced, re-processed 408\,\textrm{MHz} Haslam map \citep{1981A&A...100..209H, 1982A&AS...47....1H, 2015MNRAS.451.4311R} as a template and scaled using a spectral index and curvature model. This corresponds to the \textsc{PySM} s3 model. Pixel values at $4.76\,\mathrm{GHz}$ range from $\sim 3000\,\mu\mathrm{K}$ to $\sim 8\times 10^{5}\,\mu\mathrm{K}$, with a mean value of $\sim 2\times 10^{4}\,\mu\mathrm{K}$.
    \item \textbf{Free-Free Emission}  This was generated with the analytic model used in the \textit{Planck} \textsc{Commander} analysis \citep{2011piim.book.....D, 2016A&A...594A..10P, 2016A&A...594A...9P}. The free-free template is scaled using a single power law. This corresponds to the \textsc{PySM} f1 model. Pixel values at $4.76\,\mathrm{GHz}$ range from $\sim 90\,\mu\mathrm{K}$ to $\sim 5\times 10^{6}\,\mu\mathrm{K}$, with a mean value of $\sim 1\times 10^{4}\,\mu\mathrm{K}$.
    \item \textbf{Anomalous Microwave Emission (AME)}   This was generated by modelling emission caused by the sum of two spinning-dust populations. Scaling is calculated using the \textsc{SpDust2} code \citep{2009MNRAS.395.1055A}. This corresponds to the \textsc{PySM} a2 model. Pixel values at $4.76\,\mathrm{GHz}$ range from $\sim 5\,\mu\mathrm{K}$ to $\sim 2\times 10^{5}\,\mu\mathrm{K}$, with a mean value of $\sim 300\,\mu\mathrm{K}$.
    \item \textbf{Thermal Dust}  This was generated by modelling the emission caused by a single modified black-body. This corresponds to the \textsc{PySM} d1 model \citep{2016A&A...594A..10P}. Pixel values at $4.76\,\mathrm{GHz}$ range from $\sim 0.01\,\mu\mathrm{K}$ to $\sim 20\,\mu\mathrm{K}$, with a mean value of $\sim 0.4\,\mu\mathrm{K}$.
    \item \textbf{CMB} A lensed CMB realization was generated using the Taylens code \citep{2013JCAP...09..001N}, as part of \textsc{PySM}. This corresponds to the \textsc{PySM} c1 model. Pixel values at $4.76\,\mathrm{GHz}$ range from $\sim -300\,\mu\mathrm{K}$ to $\sim 300\,\mu\mathrm{K}$, with a mean value of $\sim 0.1\,\mu\mathrm{K}$.
\end{enumerate}

One of the simulated skies is shown in Fig. \ref{fig:sky sims}, along with the same sky filtered at the scale $R=1$. After running our detection algorithm over the simulated skies, we compare the detected sources to our input catalogues. From this, we calculate the completeness level of our detected sources, averaged over all the simulations, and record the absolute deviations of detected sources from their true positions. The validation results are presented in Section \ref{subsec: val comp and positions}, produced using an initial SNR threshold of 2.5 to tag candidate sources on the first loop over the sky, and retaining sources with $\mathrm{SNR}\geq 3.5$ after looping over the candidate sources. These thresholds were selected to give a catalogue reliability $\gtrsim 90$ per cent away from the Galactic plane. Here we define catalogue completeness as the fraction of sources above some given flux-density threshold that are recovered by the detection algorithm. The catalogue reliability is defined as the fraction of detected sources that are matched with at least one source in the input catalogue. 

\subsection{Catalogue Matching: Likelihood Ratios}\label{subsec: likelihood ratios}

The simulated validation catalogue is produced for a map with a resolution of $45\,\mathrm{arcmin}$. Given the large number of sources used for our input catalogue, source confusion presents a major complication when trying to match the output catalogues with the corresponding input catalogues from our Monte Carlo simulations. One source in the output catalogue will likely result from the blending of multiple sources in the input catalogue, and a single source in the input catalogue can contribute to more than one source in the output catalogue. We face identical issues when matching the real C-BASS catalogue to higher resolution catalogues such as GB6 and PMN.

To find matches between our input and output catalogues we use a likelihood ratio test. This has previously been used in matching point-source catalogues, particularly when when dealing with problems from source confusion, see e.g., \cite{1975AN....296...65R, 1992MNRAS.259..413S, 1997MNRAS.289..482M, 2001ApJ...551..921R, 2011MNRAS.411..505C, 2014MNRAS.444.2870W}. The core of the likelihood ratio test consists in calculating the ratio of the probability of a match between an input and output source being a true match, to the probability of the match being a random association. The likelihood ratio is given by
\begin{equation}
    \mathrm{LR} = \frac{q(\Delta S_{f})f(r)}{2\pi r\rho(\Delta S_{f})},
\end{equation}
where $q(\Delta S_{f})$ is the probability density function (PDF) for a given match being a true match as a function of the fractional flux-density difference ($\Delta S_f$) between the two sources, $f(r)$ is the PDF for true matches as a function of the absolute positional offset between the two sources, $r$, $2\pi r$ is the PDF for random associations assuming a uniform spatial distribution of background sources, and $\rho(\Delta S_f)$ is the PDF for random associations as a function of $\Delta S_f$.

For the positional PDF of true input-output matches, we assume a Gaussian distribution over the orthogonal coordinate axes. This gives a Rayleigh distribution for the PDF of true input-output matches as a function of $r$, i.e.
\begin{equation}
    f(r) = \frac{r}{\sigma_{r}}\exp\left(-\frac{r^{2}}{2\sigma_{r}^{2}}\right),
\end{equation}
where $\sigma_{r}$ is a scaling parameter. The position and flux-density PDF parameters are estimated by comparing our output source catalogues with the corresponding input catalogues. We determine $f(r)$ by searching for all candidate matches between the input and output catalogues, using a search radius of $45\,\mathrm{arcmin}$ around each output source. Candidate matches will consist of true input-output matches, along with random background associations. Assuming a uniform distribution of background sources the random matches will follow a linear trend, scaling with the sky area enclosed by a given search radius. We therefore fit $f(r)$ plus a linear trend to the histogram of absolute positional offsets for candidate input-output matches to obtain an estimate for $f(r)$.

To determine $\rho(\Delta S_f)$, we generate randomized output catalogues by randomly distributing our output sources over the sky and randomizing source flux-densities from the parent distribution. We then find candidate matches between input and randomized output sources, again using a $45\,\mathrm{arcmin}$ search radius. We bin the candidate matches as a function of the fractional flux-density difference between the input and output source, given by
\begin{equation}
    \Delta S_f = \frac{S_{\mathrm{out}} - S_{\mathrm{in}}}{S_{\mathrm{in}}},
\end{equation}
where $S_{\mathrm{out}}$ is the output source flux-density and $S_{\mathrm{in}}$ is the input source flux-density. This gives us the number of random matches as a function of $\Delta S_f$ and hence our estimate for $\rho(\Delta S_f)$. To obtain $q(\Delta S_f)$ we carry out the same procedure as above, but with the real output catalogues. We then estimate $q(\Delta S_f)$ from the difference between the number of matches as a function of $\Delta S_f$ for the real output catalogues, and the number of matches for the randomized catalogues, i.e. from the excess matches in the real output catalogues.

To determine the $\mathrm{LR}$ threshold for declaring a true match we calculate the likelihood ratios for all candidate matches between the input catalogues and randomized output catalogues. From this we determine the value of $\mathrm{LR}$ at which we would declare $10$ per cent of our randomized matches to be true matches. This value is used as the $\mathrm{LR}$ threshold at which we declare a true input-output match. This is sufficient for limiting the fraction of false identifications with the input catalogue, whilst ensuring a low probability of spuriously identifying a source in the output catalogue as a true positive, given that matched output sources are typically associated with multiple input sources.

\subsection{Validation: Completeness, Position and Flux-Density Accuracy}\label{subsec: val comp and positions}

\begin{figure}
    \centering
    \includegraphics[width=\columnwidth]{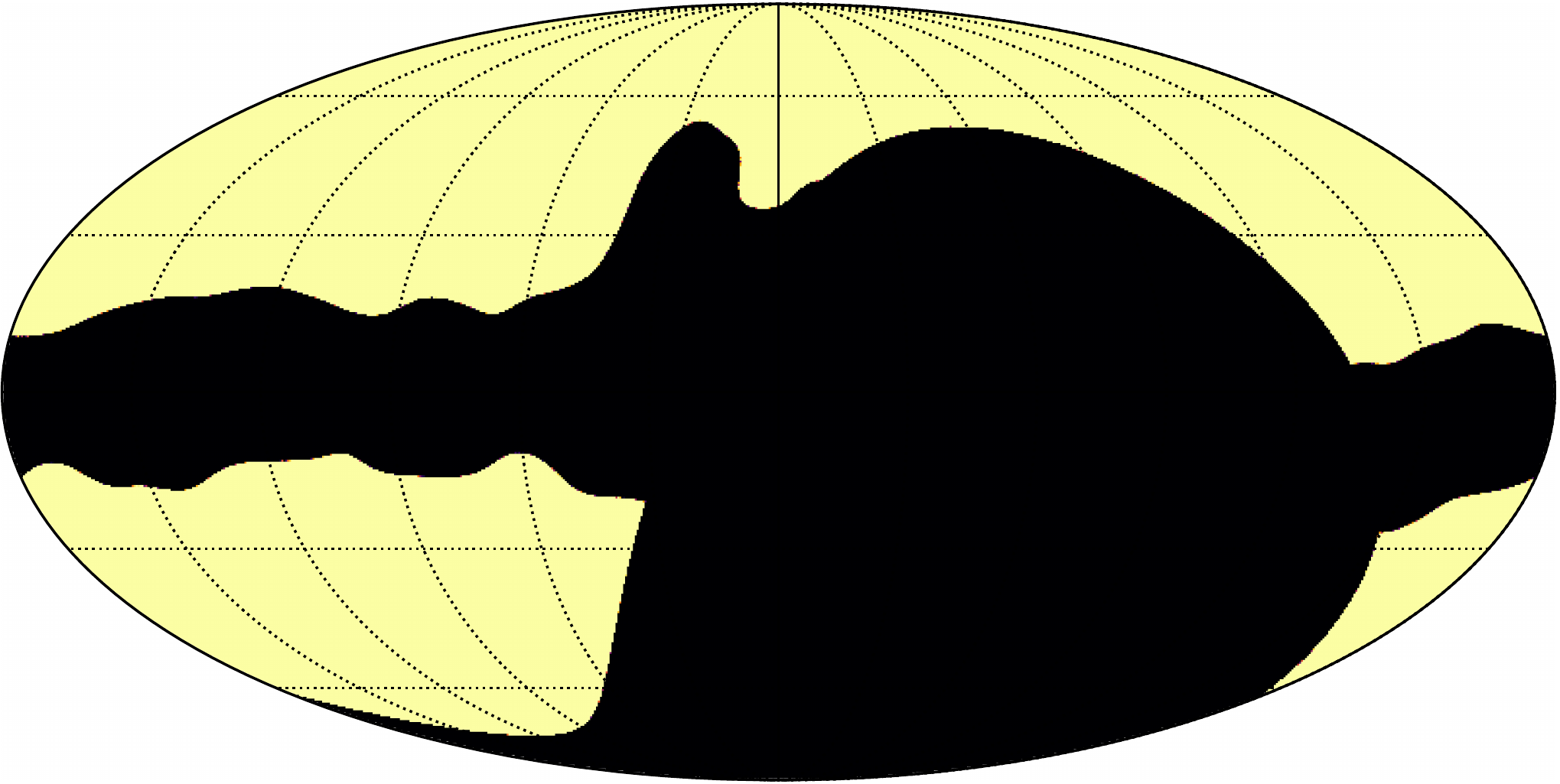}
    \caption{The combined Galactic plane and C-BASS North mask used in our analysis of the C-BASS catalogue, shown here in Galactic coordinates. The mask was constructed to cover the brightest $30$ per cent of the northern sky observed by C-BASS, along with declinations $\delta<-10^{\circ}$. We refer to this mask as the CG30 mask.}
    \label{fig: galactic masks}
\end{figure}

We obtain the catalogue completeness and reliability, averaged over our Monte Carlo simulations, to confirm the ability of the source-detection algorithm to accurately recover an input source population. Throughout our analysis we use the mask shown in Fig. \ref{fig: galactic masks} to characterize the catalogue properties away from the brightest diffuse emission along the Galactic plane. The mask was produced by smoothing the C-BASS northern sky map with a $10^{\circ}$ FWHM Gaussian and masking the brightest $30$ per cent of pixels. We also mask all pixels corresponding to declinations below $\delta<-10^{\circ}$ in order to match the sky area observed by the C-BASS northern-sky survey. We refer to this mask as the CG30 mask. 

In determining the positional offset of the detected source from the input sources, we compare its position to the photocentre of matched input sources. That is, after matching to input sources using the $\mathrm{LR}$ test, we determine the weighted average position of the matched input sources, using the input source flux-densities as weights. In comparing flux densities, we sum flux densities for matched sources, weighted according to the C-BASS beam profile, i.e.
\begin{equation}
S_{\nu}^{\mathrm{match}}=\sum_{i}S_{\nu}^{\left(i\right)}\mathcal{B}(|\mathbfit{x}_{i}-\mathbfit{x}_{\mathrm{det}}|),
\label{weighted flux}
\end{equation}
where the sum runs over matched sources in the input catalogues, $S_{\nu}^{\left(i\right)}$ is the flux-density of the $i^{\mathrm{th}}$ matched source, $\mathbfit{x}_{i}$ is the position of the $i^{\mathrm{th}}$ matched source, $\mathbfit{x}_{\mathrm{det}}$ is the position of the detected source and $\mathcal{B}$ is the C-BASS beam profile.

\begin{figure*}
    \centering
    \subfigure[Completeness map]{\includegraphics[width=\columnwidth]{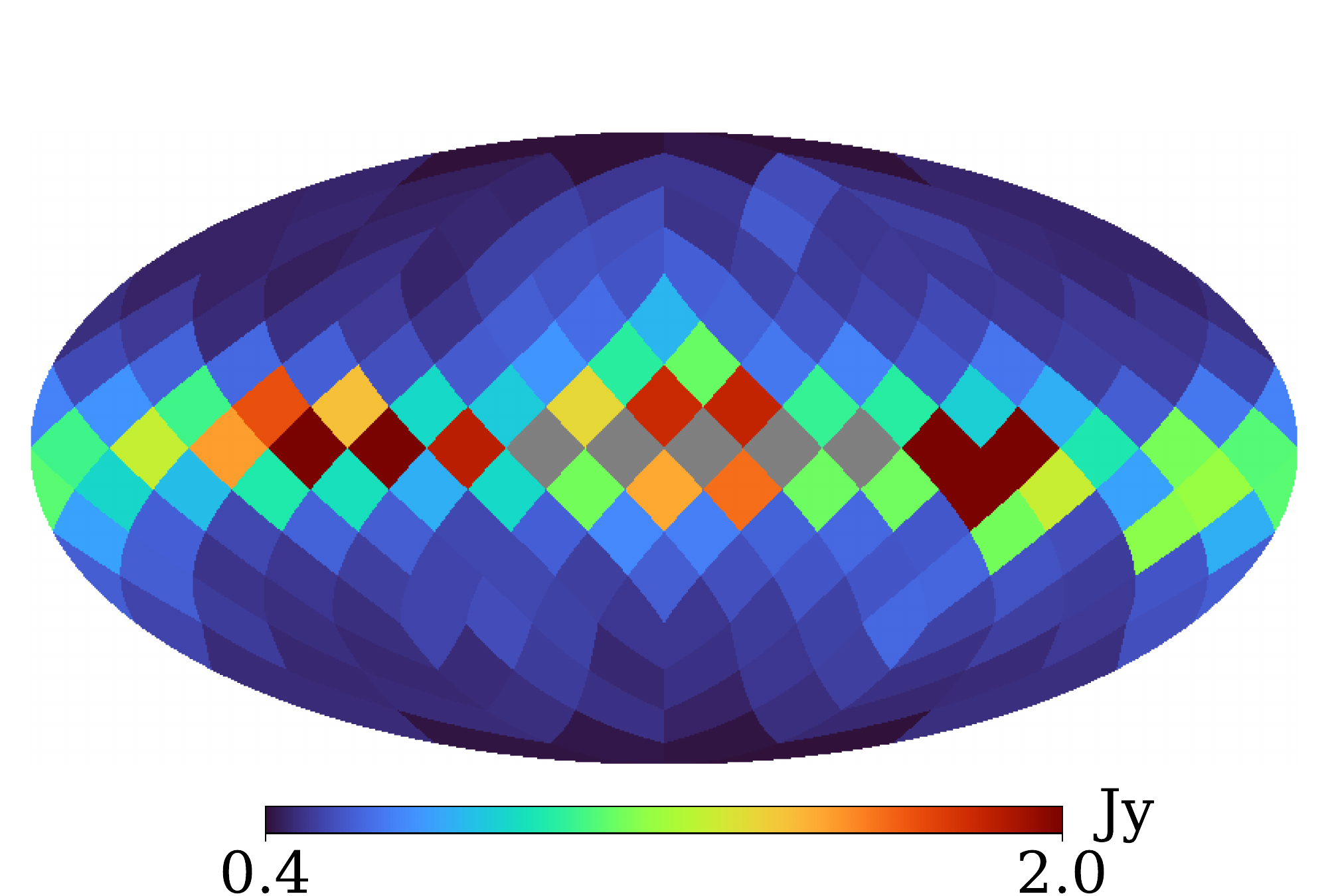}}
    \subfigure[Reliability map]{\includegraphics[width=\columnwidth]{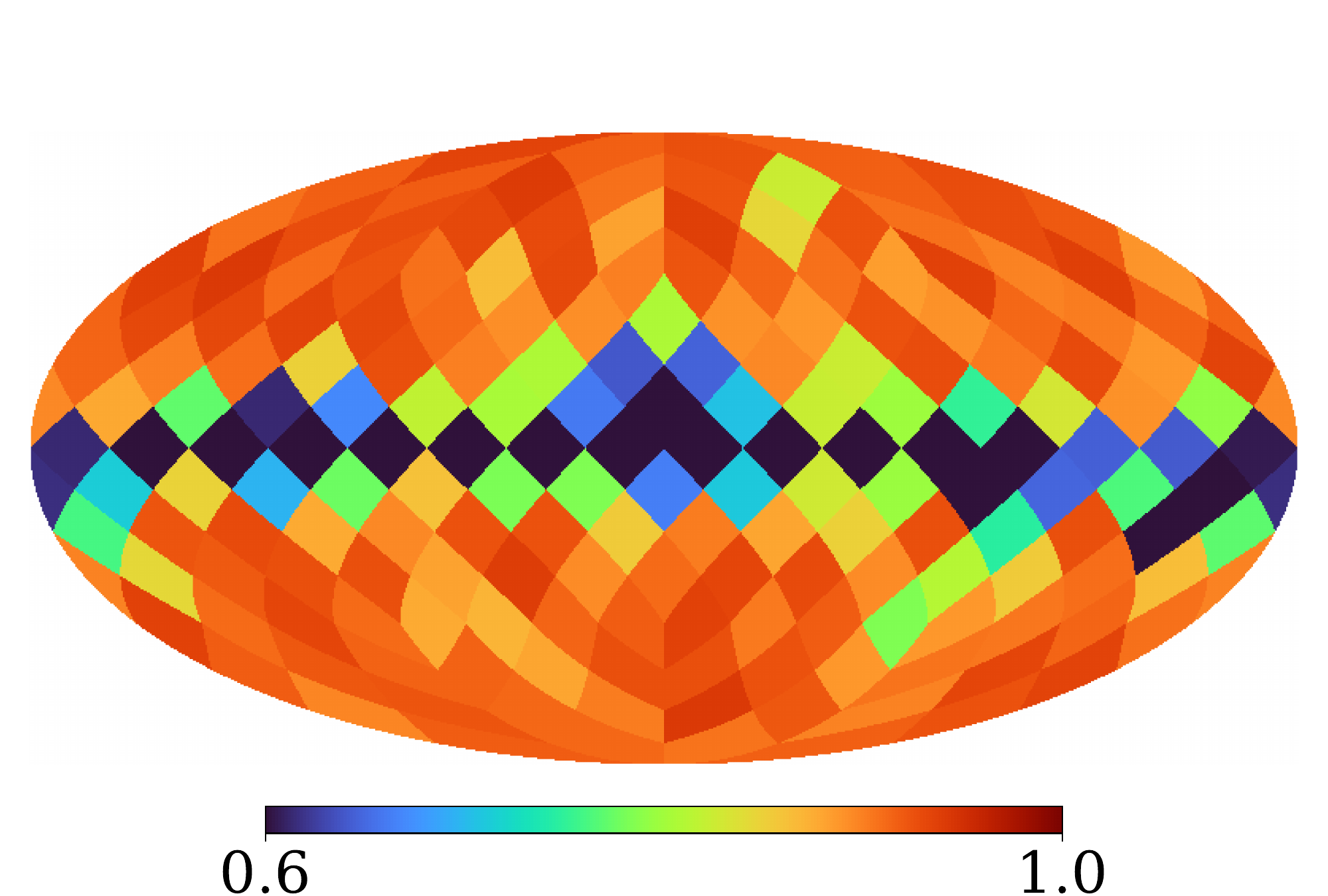}}
    \caption{Panel (a): A map of the completeness estimates obtained from the validation simulations. Panel (b): A map of the corresponding reliability estimates. These estimates were calculated on \textsc{NSIDE}=4 sky pixels and are shown in Galactic coordinates. In pixels outside of the CG30 mask, the mean 90 per cent completeness level is approximately $500\,\mathrm{mJy}$, with a corresponding mean reliability of $91$ per cent. It can be seen that along the Galactic plane the completeness and reliability is significantly lower than at higher Galactic latitudes. This is to be expected given the stronger diffuse emission that is not removed by the SMHW2 filter. Grey pixels in the completeness map denote pixels where $90$ per cent completeness was never achieved.}
    \label{fig: val comp}
\end{figure*}

In Fig. \ref{fig: val comp} we show maps of the completeness and reliability in NSIDE=4 pixels obtained for the validation simulations. We can see that close to the Galactic plane the 90 per cent completeness level is higher and the catalogue reliability is lower. Indeed, for some pixels along the Galactic plane 90 per cent completeness is never achieved. This is to be expected given the increased intensity of diffuse emission along the Galactic plane, which is not fully removed by the SMHW2 filter. Over regions of the sky left un-masked by the CG30 mask, we find a mean 90 per cent catalogue completeness of approximately $500\,\mathrm{mJy}$, corresponding to a mean reliability of approximately $91$ per cent.

\begin{figure}
    \centering
    \includegraphics[width=\columnwidth]{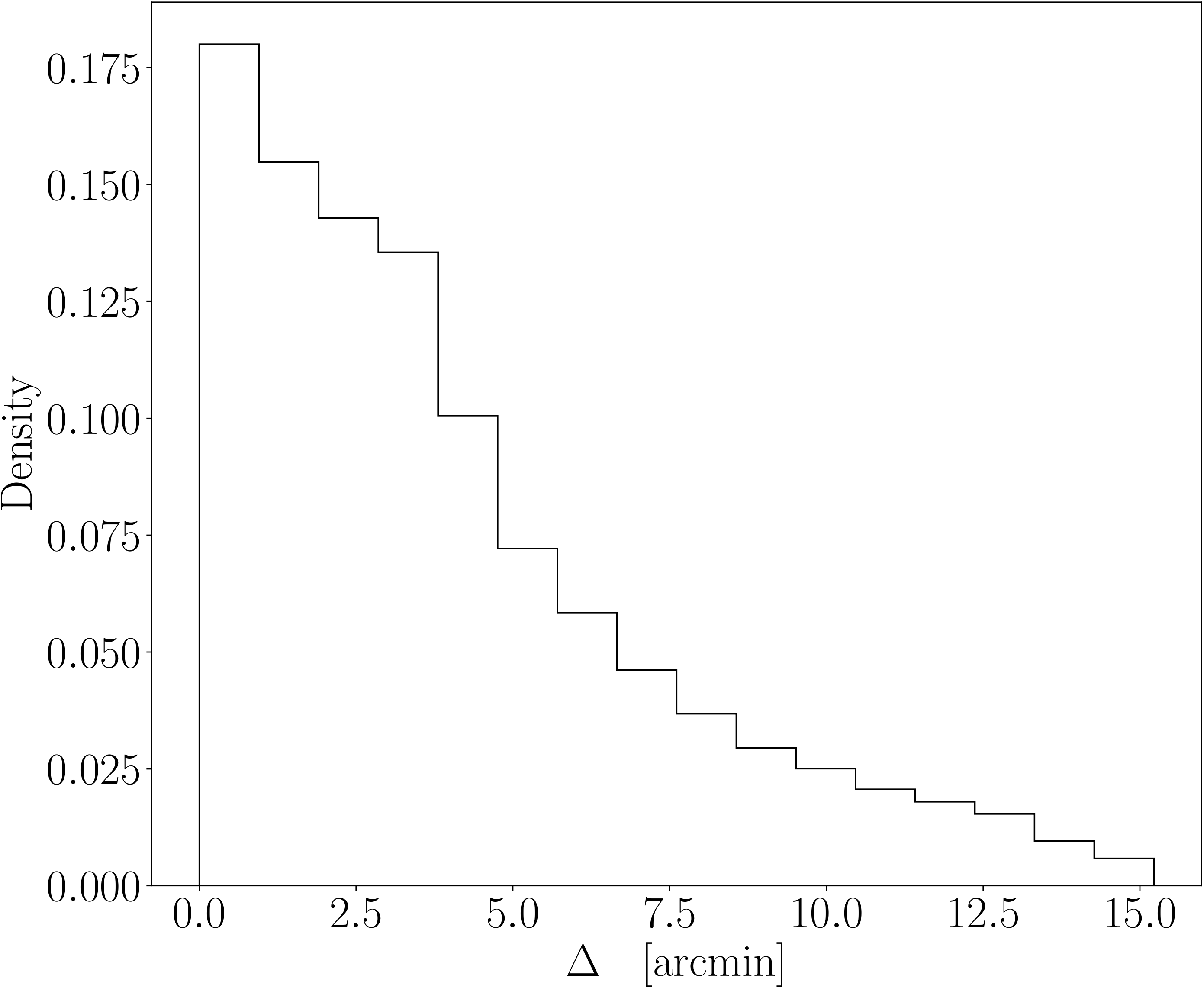}
    \caption{The histogram of measured absolute positional offsets, $\Delta$, applying the CG30 mask, obtained for the 100 Monte Carlo simulations. Over the un-masked sky area we find the detected positional offsets peak around zero, with a median positional offset of $3.2\,\mathrm{arcmin}$, which is small compared to the C-BASS beam of $45\,\mathrm{arcmin}$. Larger positional offsets are predominantly driven by fainter detected sources below $\sim 1\,\mathrm{Jy}$, and sources detected in regions of brighter diffuse emission. The positional offsets of all the detected sources will consist of samples drawn from Rayleigh distributions, which results in a tail of larger offsets beyond the peak of the positional offset distribution.
    }
    \label{fig: val pos offset}
\end{figure}

From the validation simulations, we calculate the absolute positional offsets of detected sources from input sources. These cluster below 10\,\textrm{arcmin}, peaking around zero, with a median offset of approximately $3.2\,\mathrm{arcmin}$. Considering the $45\,\mathrm{arcmin}$ resolution of the maps used for source detection, and the significant confusion effects at this resolution, this is a small positional uncertainty for our source detection algorithm. Indeed, for the purposes of producing a point-source mask for the actual C-BASS maps this is more than sufficient. The histogram of absolute positional offsets of detected sources, applying the CG30 mask, is shown in Fig. \ref{fig: val pos offset}.

\begin{figure*}
    \centering
    \subfigure[Flux-Flux plot]{\includegraphics[width=\columnwidth]{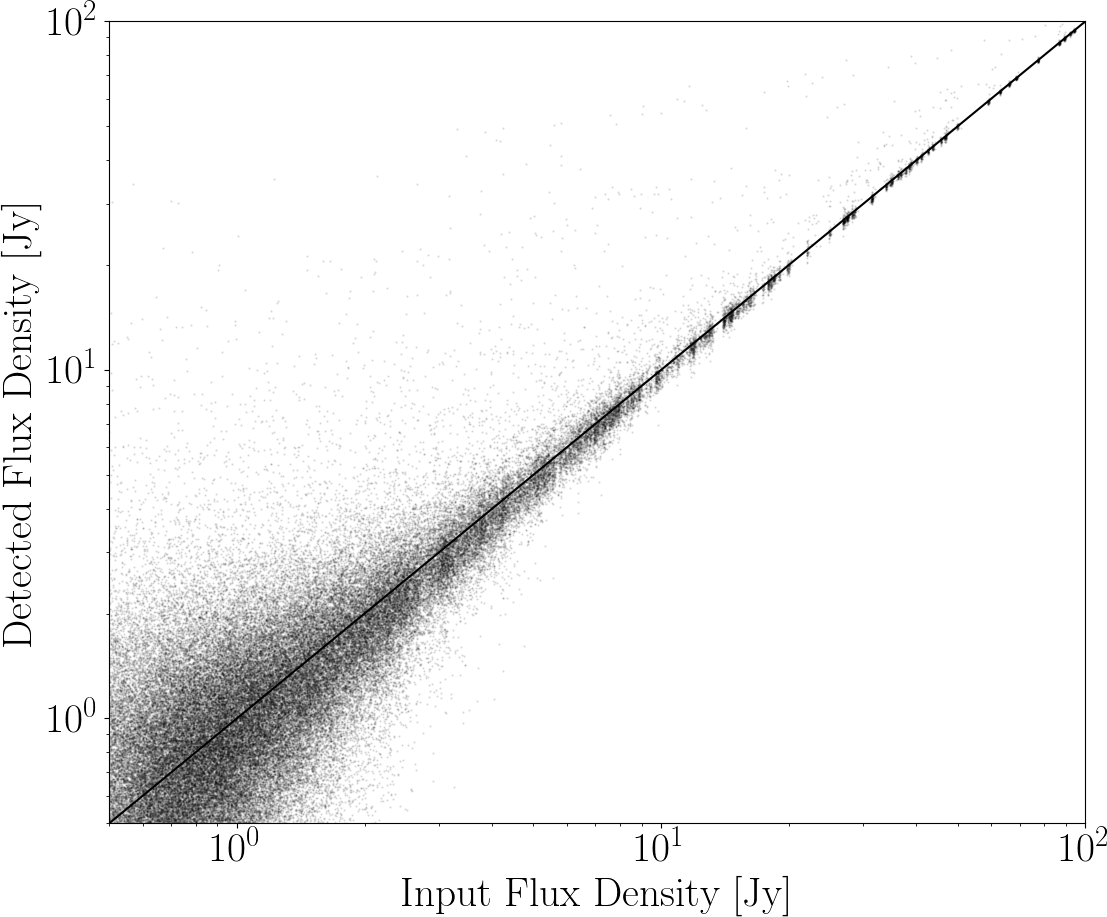}}
    \subfigure[Flux-density ratio histogram]{\includegraphics[width=\columnwidth]{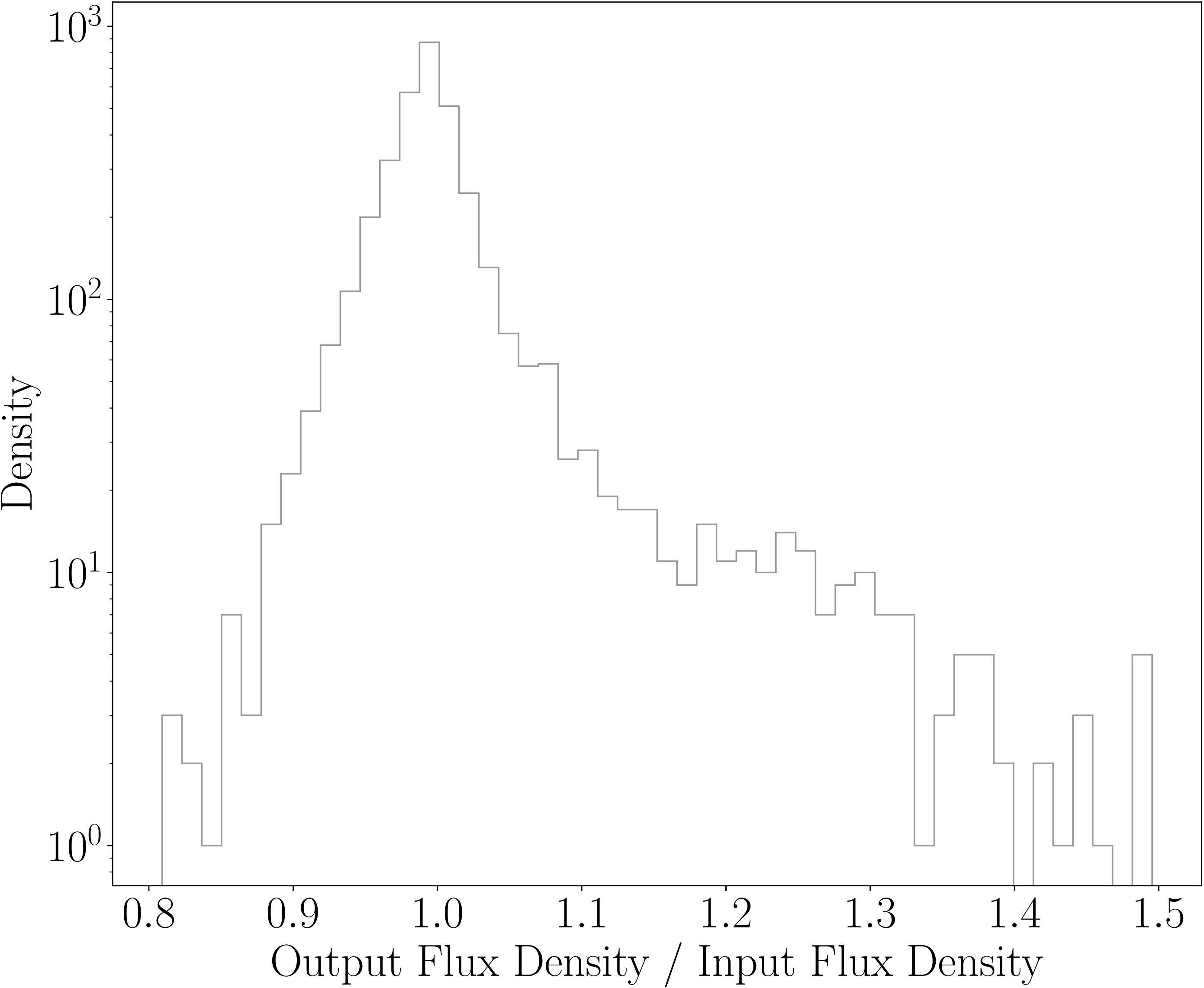}}
    \caption{Panel (a): The detected source flux densities against the matched input flux densities for the 100 Monte Carlo simulations, applying the CG30 mask. Panel (b): The histogram of the detected flux densities, divided by the matched input flux-densities, for sources with input flux densities greater than $10\,\mathrm{Jy}$. We can see that for the majority of bright sources we obtain reliable flux-density estimates. However, there is a small population ($\sim 1$ per cent) of detected sources with spuriously high flux-density estimates, due to confusion effects.}
    \label{fig: val flux match}
\end{figure*}

We are able to recover input source flux-densities using aperture photometry, with a plot of detected flux densities against input flux densities being shown in Fig. \ref{fig: val flux match}. At flux densities greater than $1\,\mathrm{Jy}$ we recover reasonable flux-density estimates for the majority of sources. A small fraction of sources are recovered with flux densities biased slightly high. This is to be expected given the $45\,\mathrm{arcmin}$ resolution of the C-BASS map, meaning that multiple input sources are associated with each detected source. The detected source flux density will be the summed contribution from associated input sources. For flux densities below $\sim 1\,\mathrm{Jy}$ the scatter in the recovered flux densities reaches the same magnitude as the input flux densities. This is again to be expected; below this level sources become much more heavily obscured by surrounding diffuse emission, making photometric estimates of source flux densities significantly more unreliable. We note here that there is a small cluster of recovered sources at these low flux densities where the recovered flux-density estimate is spuriously high. This is largely caused by sources being detected in regions of bright diffuse and highly extended emission, leading to errors in the background estimation. 

Our validation simulations have demonstrated that our source-detection algorithm is able to accurately recover an underlying source population, from a sky consisting of additional contributions from diffuse emission and noise. 
\section{The C-BASS Northern Sky Catalogue}\label{sec: cbass catalogue}

In this section we present the C-BASS northern sky, total-intensity point-source catalogue. The point-source catalogue was produced using data obtained from only night-time observations, combining data from the various elevation scans of the telescope. C-BASS data were calibrated against Tau A as a primary calibrator and Cas A as a secondary calibrator, with calibrator flux densities being calculated from the spectral forms in \cite{2011ApJS..192...19W}. For Tau A the spectral model takes the form
\begin{equation}
    \log S_{\nu}(\mathrm{Jy}) = 2.506-0.302\log\left(\frac{\nu}{40\,\mathrm{GHz}}\right),
\end{equation}
and for Cas A it takes the form
\begin{equation}
    \log S_{\nu}(\mathrm{Jy}) = 2.204-0.682\log\left(\frac{\nu}{40\,\mathrm{GHz}}\right)+0.038\log^{2}\left(\frac{\nu}{40\,\mathrm{GHz}}\right).
\end{equation}
These spectral models give the flux density of Tau A at $4.76\,\mathrm{GHz}$ as $S_{\nu}=609.8\pm 4.2\,\mathrm{Jy}$ in epoch 2005, and the flux-density of Cas A as $S_{\nu}=736.1\pm 3.4\,\mathrm{Jy}$ in epoch 2000. The secular decreases adopted for Tau A and Cas A were the same as those used for the model fits in \cite{2011ApJS..192...19W}. For Tau A this was -0.167 per cent per year, which was taken from \cite{2010ApJ...711..417M}, and for Cas A the secular decrease was -0.53 per cent per year. The C-BASS calibration is accurate to $\sim 5$ per cent. The flux-temperature conversion is defined through
\begin{equation}
    T_{\mathrm{source}} = S_{\nu}(\mathrm{Jy})\frac{\pi}{2k_{B}\times10^{26}}\left(\frac{6.1}{2}\right)^{2}\alpha_{\mathrm{eff}},
\end{equation}
where $T_{\mathrm{source}}$ is the source temperature, $k_{B}$ is the Boltzmann constant and $\alpha_{\mathrm{eff}}=0.55$ is the theoretical aperture efficiency for C-BASS. Detailed discussions of the C-BASS data reduction and calibration will be given in the C-BASS survey and commissioning papers (Taylor et al., in prep. and Pearson et al., in prep). We perform the source detection on the C-BASS map at NSIDE=1024, using the high NSIDE map to allow for the precise determination of source positions.   

The outline of the remainder of this section is as follows: In Section \ref{subsec: cbass cat summary} we summarize the catalogue properties. In Section \ref{subsec: gb6 matching} we present the results from matching the C-BASS and GB6/PMN catalogues, using these results to quantify the C-BASS pointing accuracy and flux-density scale. In Section \ref{subsec: outliers} we comment briefly on sources detected with outlier flux densities. In Section \ref{subsec: diff source counts} we show the differential source counts for the C-BASS catalogue, and compare these to the GB6 source counts. Additional comparisons were also made with sources in the RATAN-600 catalogue over the NCP region. In this case we recovered the expected bright source population, with the results obtained through these comparisons being consistent with those from comparisons with the GB6/PMN catalogues.

\subsection{Northern Sky Intensity Catalogue: Summary}\label{subsec: cbass cat summary}

\begin{figure*}
\centering
\includegraphics[width=\textwidth]{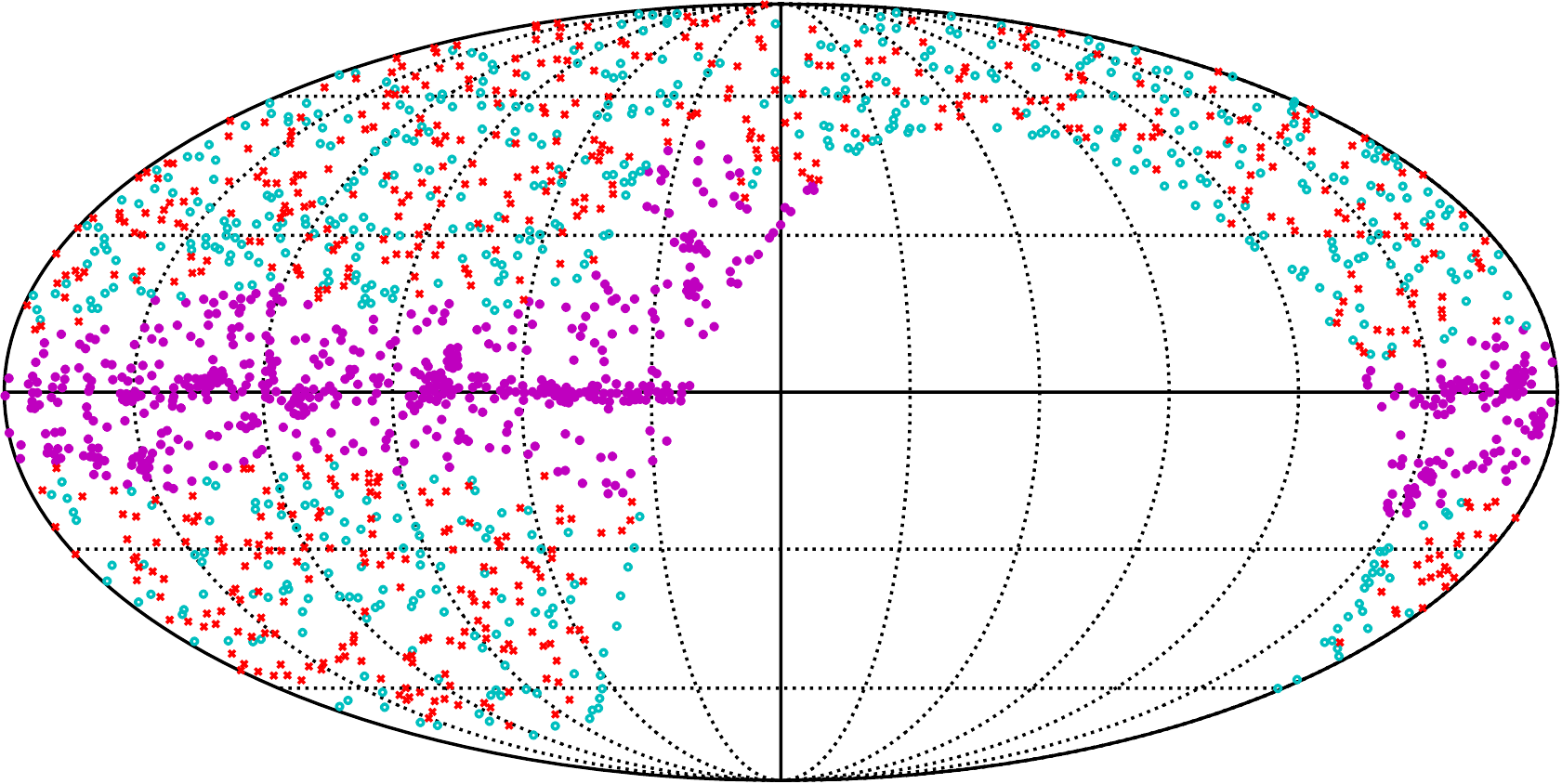}
\caption{Positions of the northern sky, total intensity C-BASS point sources. Positions shown here are in Galactic coordinates. \textit{Filled purple circles} denote sources that fall within the CG30 mask. \textit{Open cyan circles} denote sources outside the mask, with a fitted flux density below the $90$ per cent completeness level ($610\,\mathrm{mJy}$), and \textit{red crosses} denote the remaining sources with fitted flux densities above the $90$ per cent completeness level.}
\label{fig: cbass source positions}
\end{figure*}

For our default catalogue, we use the same SNR thresholds as for the validation analysis i.e., $\mathrm{SNR\geq2.5}$ to tag candidate sources on the first loop over the sky, followed by a final detection threshold of $\mathrm{SNR\geq3.5}$ after looping over the candidate source locations. To estimate the completeness and reliability of the real C-BASS catalogue we compare it with the GB6 and PMN catalogues over their areas of common sky coverage. Applying the CG30 mask we obtain a catalogue reliability of approximately $98$ per cent, with a corresponding $90$ per cent completeness level of $610\,\mathrm{mJy}$, recovering the expected bright source population at $4.76\,\mathrm{GHz}$. The confusion level in the C-BASS map is approximately $85\,\mathrm{mJy}$ \citep{2018MNRAS.480.3224J}. Below this level sources cannot be individually detected in our maps, acting as an additional noise contribution. Our analysis has resulted in a catalogue of 1784 point sources covering declinations $\delta\geq-10^{\circ}$. The catalogue properties are summarized in Table \ref{tab: cbass summary}, with the detected source positions for the default catalogue shown in Fig. \ref{fig: cbass source positions}.

Beam sidelobes around bright sources can result in spurious source detections. The first sidelobe of the C-BASS beam has some azimuthal structure, and has a peak value of about $2.5$ percent of the peak. Given the source detection level of around 500~mJy, we might expect spurious detections from sources brighter than 20~Jy. Outside of the CG30 mask there are three sources brighter than this, 3C273, Virgo A and 3C279.  Therefore we do not allow any additional detections within a $2^{\circ}$ radius exclusion zone centered on these three sources. In constructing a point-source mask, any issues from spurious sidelobe detections can be further mitigated by masking larger areas around the brightest sources.

It is important to note here that, whilst there are a large number of source detections in the Galactic plane, the reliability of these detections is significantly lower than at higher Galactic latitudes. This is a result of the more intense diffuse emission obscuring point sources, and highly extended emission in the Galaxy being spuriously identified as point sources. For the purposes of scientific analyses with other CMB datasets this is not a major issue, given that much of the Galactic plane will be masked out for such analyses. 

\begin{table}
\caption{C-BASS northern sky, total intensity catalogue summary. Position uncertainties include intrinsic uncertainty from the source detection algorithm, along with additional errors from the C-BASS pointing.}
\label{tab: cbass summary}
\begin{threeparttable}
\begin{tabular}{lcc}
\hline
& C-BASS Catalogue\\
\hline
Frequency \dotfill & $4.76\,\mathrm{GHz}$\\
Map FWHM\dotfill & $45\,\mathrm{arcmin}$ \\
SNR threshold\dotfill & 3.5\\
Reliability: &\\
\hspace{3mm} Full Sky\dotfill & 90\%\\
\hspace{3mm} CG30 mask\dotfill & 98\%\\
Number of sources: &\\
\hspace{3mm} Full sky\dotfill & 1784\\
\hspace{3mm} CG30 mask\dotfill & 1136\\
90\% completeness: & \\
\hspace{3mm} Full sky\dotfill & $1000\,\mathrm{mJy}$\\
\hspace{3mm} CG30 mask\dotfill & $610\,\mathrm{mJy}$\\
$N\left(S>S_{90}\right)^{a}$: & \\
\hspace{3mm} Full sky\dotfill & 793\\
\hspace{3mm} CG30 mask\dotfill & 560\\
Position uncertainty &\\
\hspace{3mm} Full sky\dotfill & $3.4\,\mathrm{arcmin}$\\
\hspace{3mm} CG30 mask\dotfill & $3.2\,\mathrm{arcmin}$\\
\hline
\end{tabular}
\begin{tablenotes}
\item $^a$ This is the number of sources with measured flux densities greater than the $90$ per cent completeness level.
\end{tablenotes}
\end{threeparttable}

\end{table}

\subsection{GB6 and PMN Matching}\label{subsec: gb6 matching}

To cross-check the C-BASS pointing and flux-density scale, we match the C-BASS point sources with the GB6 and PMN catalogues over their areas of common sky coverage. Similar to the situation encountered during the algorithm validation, the GB6 and PMN catalogues were produced using much higher resolution surveys than C-BASS. What may appear as a single source in the C-BASS maps could actually correspond to multiple sources in the GB6 and PMN catalogues. Matches between catalogues are found using the likelihood ratio technique described in Section \ref{subsec: likelihood ratios}. In characterizing the C-BASS pointing, we compare the C-BASS source position with the photocentre (defined in Section \ref{subsec: likelihood ratios}) of matched GB6 or PMN sources. In comparing flux densities we use the beam-weighted sum of matched source flux densities, defined in Equation \ref{weighted flux}.

\begin{figure}
\centering
\includegraphics[width=\columnwidth]{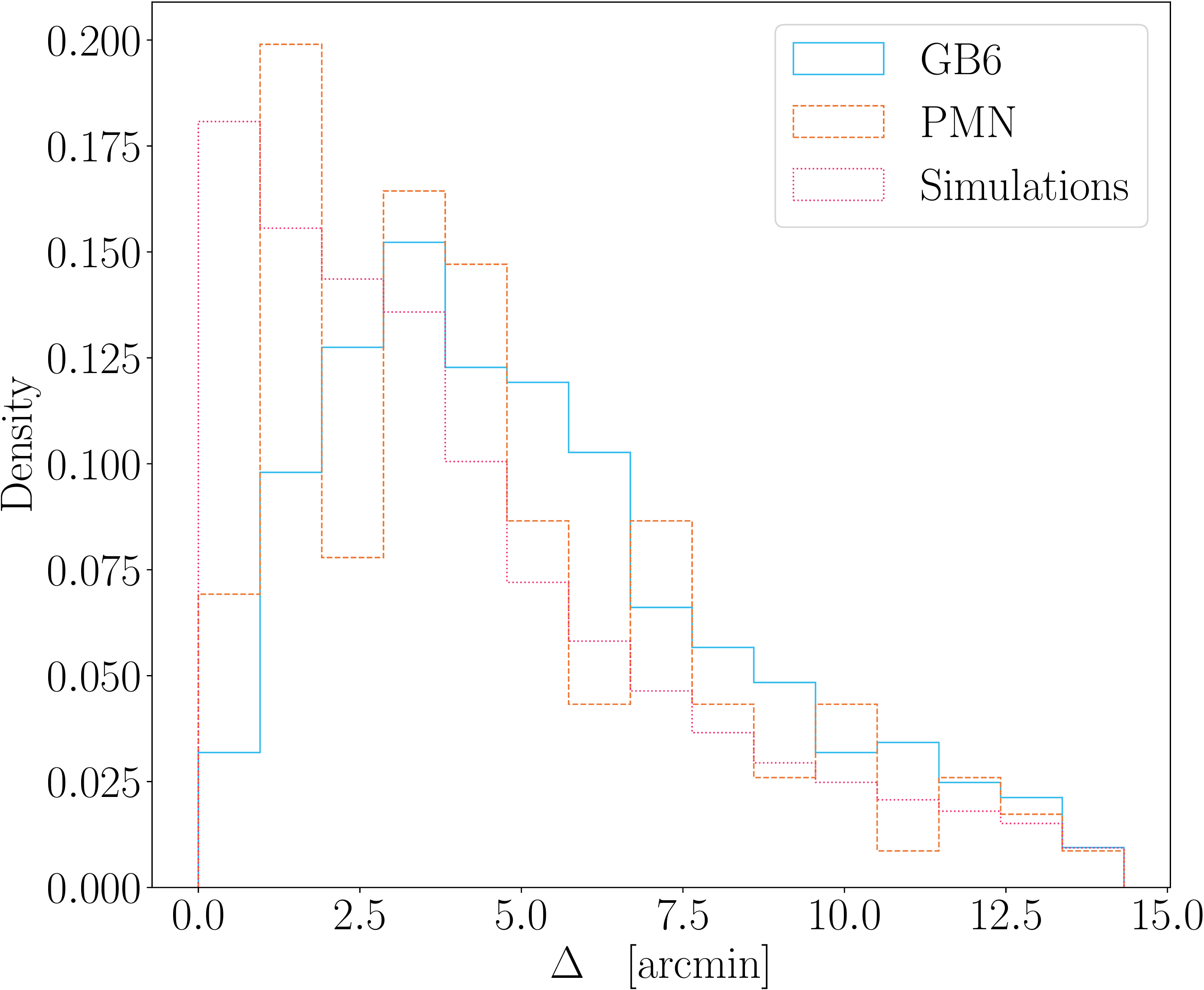}
\caption{A histogram of the absolute positional offsets of C-BASS sources from matched GB6 and PMN sources, applying the CG30 mask, plotted alongside the corresponding histogram of positional offsets obtained from the validation simulations. We use the photocentres of matched GB6 and PMN sources for comparison. Positional offsets cluster below $10\,\mathrm{arcmin}$, peaking at approximately $3.2\,\mathrm{arcmin}$. We can see that the peak of the positional offsets obtained for the real C-BASS catalogue peak $\sim 3\,\mathrm{arcmin}$ higher than for the validation simulations. This is to be expected given there is some level of pointing jitter in the real C-BASS data that is not present in the simulations. We obtain similar distributions for the absolute positional offsets with both GB6 and PMN.}
\label{fig: gb6 delta}
\end{figure}

In Fig. \ref{fig: gb6 delta} we show a histogram of the absolute positional offsets of C-BASS sources from matched GB6 and PMN sources, using the photocentres of matched GB6 and PMN sources for comparison. The offsets are clustered below $10\,\mathrm{arcmin}$, peaking at approximately $3.2\,\mathrm{arcmin}$\footnote{Note that the position of the mode for a Rayleigh distribution is given by $\sigma_{r}$.}. This is broadly consistent with the results obtained from our Monte Carlo simulations during the algorithm validation, with additional errors arising here from the C-BASS pointing. The distributions of the absolute positional offsets when matching to GB6 and PMN are also very similar. Larger positional offsets ($\gtrsim 10\,\mathrm{arcmin}$) are predominantly driven by fainter detected sources below $\sim 1\,\mathrm{Jy}$ and sources detected in regions of brighter diffuse emission. Further, given a set of positional offsets drawn from Rayleigh distributions, we expect the overall distribution to have a tail of larger offsets. The typical positional offsets determined using the C-BASS source catalogue are consistent with the estimated pointing accuracy of C-BASS (Pearson et al. (in prep.)). 

\begin{figure*}
\centering
\subfigure[No Galactic mask]{\includegraphics[width=0.48\textwidth]{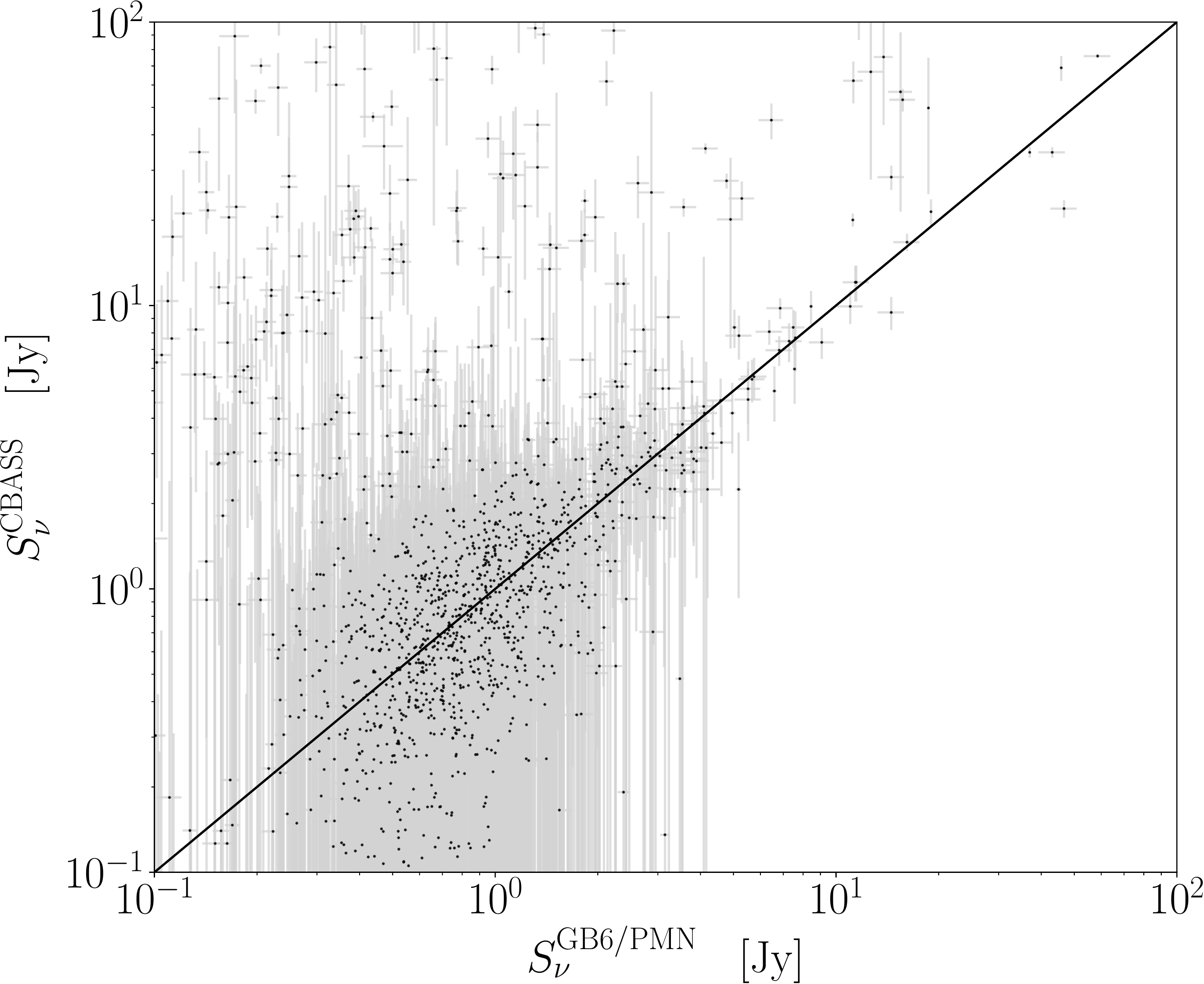}}
\subfigure[CG30 Mask]{\includegraphics[width=0.48\textwidth]{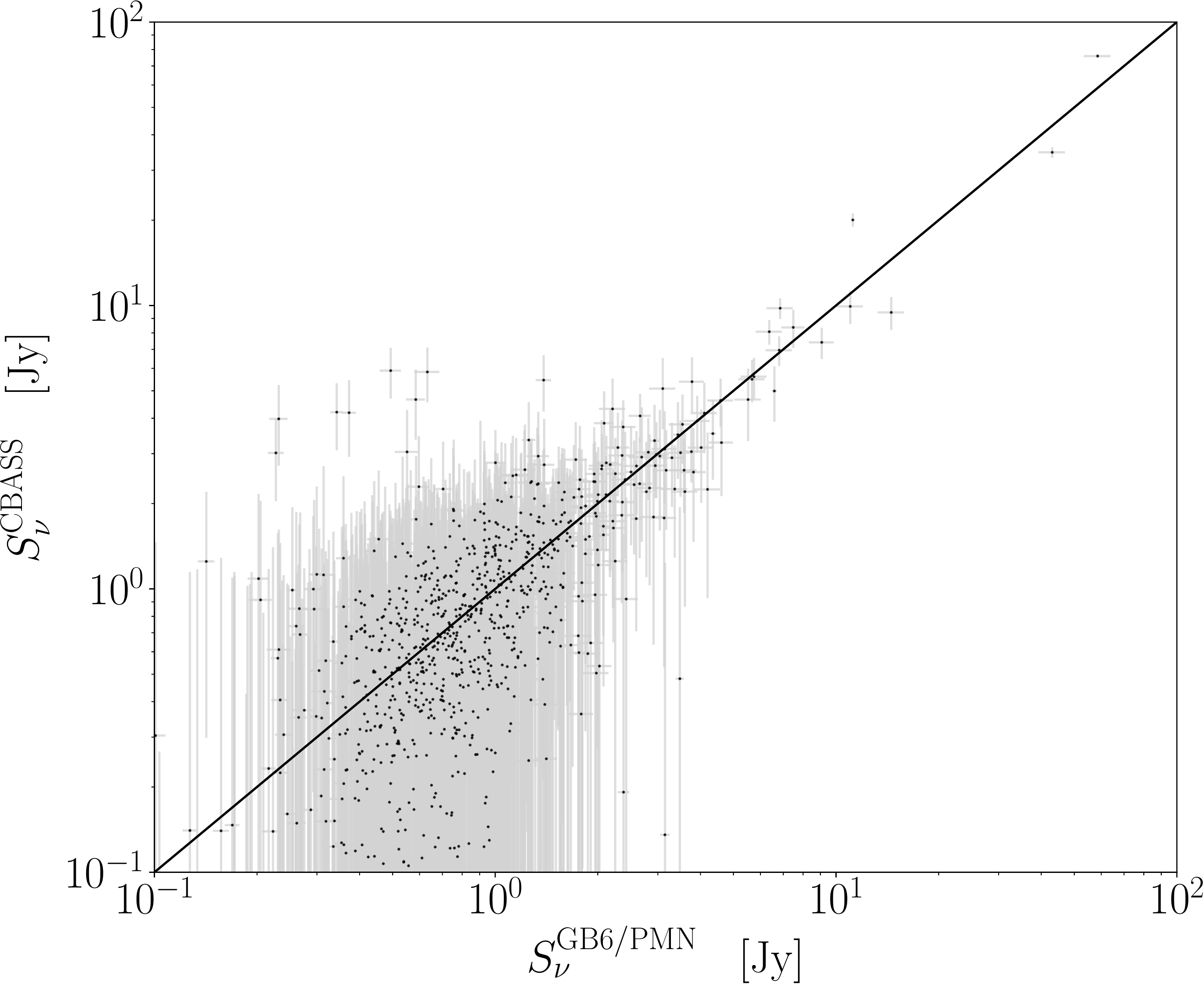}}
\caption{Flux-flux plots for matched C-BASS and GB6/PMN sources. Panel (a): A plot of C-BASS flux densities against matched GB6/PMN flux densities, with no mask applied. Panel (b): A plot of C-BASS flux densities against matched GB6/PMN flux densities, applying the CG30 mask. We also plot the line $S_{\nu}^{\mathrm{GB6/PMN}}=S_{\nu}^{\mathrm{CBASS}}$ in each panel. When we do not apply a Galactic mask, we can see a group of C-BASS sources with significantly higher flux densities than the matched GB6/PMN sources. This can be attributed to errors in fitting for source flux densities in the Galactic plane, particularly in the presence of bright highly extended emission. On applying a Galactic plane mask, this group of sources is largely removed and sources cluster more tightly around $S_{\nu}^{\mathrm{GB6/PMN}}=S_{\nu}^{\mathrm{CBASS}}$. Apparent outliers in the flux-density matches can largely be attributed to variable, flat-spectrum sources. Removing any C-BASS sources identified with flat-spectrum sources in the CRATES catalogue, we remove most of these outliers and improve the consistency between the C-BASS and GB6/PMN flux-density scales.}
\label{fig: gb6 cbass flux}
\end{figure*}

We also plot the C-BASS source flux densities (determined using aperture photometry, as described in Section \ref{subsec: algorithm structure}) against matched GB6 and PMN flux-densities, using Equation \ref{weighted flux} to determine the matched flux density, in Fig. \ref{fig: gb6 cbass flux}. When no Galactic plane mask is applied, we can see that most sources cluster around the line $S_{\nu}^{\mathrm{GB6}}=S_{\nu}^{\mathrm{CBASS}}$ ($S_{\nu}^{\mathrm{PMN}}=S_{\nu}^{\mathrm{CBASS}}$). However, there is an apparent population of sources where the measured C-BASS flux density is significantly greater than the GB6 and PMN flux densities. This is largely due to the tagging of sources in the Galactic plane, where the additional presence of very bright, highly extended emission leads to spuriously high flux-density estimates. On applying a Galactic plane mask this population is largely removed. The small number of remaining sources where C-BASS significantly over-estimates the flux densities may either result from sources sitting in regions of bright diffuse emission (e.g., within a Galactic spur) or from heavily confused sources in the C-BASS map. Most of the sources recovered with spuriously high densities are associated with low flux density GB6/PMN sources ($\lesssim 1\,\mathrm{Jy}$). At these flux densities the uncertainties in our flux-density estimates from aperture photometry become much larger, consistent with the larger spread in C-BASS to matched flux densities at these levels. 

To quantify the agreement between the C-BASS and GB6/PMN flux-density scales we fit a straight line between the C-BASS and GB6/PMN flux-densities,
\begin{equation}
    S_{\nu}^{\mathrm{CBASS}} = b S_{\nu}^{\mathrm{GB6/PMN}},
\end{equation}
where $b$ is the gradient between the C-BASS and GB6/PMN flux-densities. We use a measurement-error model \citep{stan_development_team_stan_2012} to account for uncertainties in the C-BASS and GB6/PMN flux densities by treating the true value of the GB6/PMN flux densities as missing data and marginalising over the distribution of the measured GB6/PMN flux densities. We also model the outlier distribution following \cite{2010arXiv1008.4686H}. For this we assume outliers are drawn from some broad Gaussian distribution specified by a mean $\mu_{\mathrm{out}}$ and standard deviation $\sigma_{\mathrm{out}}$. The probability of a data point being an outlier is given by the parameter $P_{\mathrm{out}}$. The full set of priors is given by,
\begin{align}
    &b\sim\mathcal{N}(\mu=1, \sigma=0.25),\\
    &\mu_{\mathrm{out}}\sim\mathcal{N}(\mu=1, \sigma=100),\\
    &\sigma_{\mathrm{out}}\sim\mathrm{Exp}(\lambda=1/100),\\
    &P_{\mathrm{out}}\sim\mathrm{Unif}(0, 1),\\
    &S_{\nu}^{x,i}\sim\mathrm{Unif}(0, \infty),\\
    &S_{\nu}^{\mathrm{GB6/PMN},i}\sim\mathcal{N}(\mu=S_{\nu}^{x,i},\sigma=\sigma_{\mathrm{GB6/PMN}}^{i}),
\end{align}
where $S_{\nu}^{x,i}$ is the true GB6/PMN flux-density for the $i^{\mathrm{th}}$ matched source, $S_{\nu}^{\mathrm{GB6/PMN},i}$ is the measured GB6/PMN flux-density for the $i^{\mathrm{th}}$ matched source and $\sigma_{\mathrm{GB6/PMN}}^{i}$ is the corresponding uncertainty on the GB6/PMN flux density. The likelihood is given by
\begin{multline}
    \mathcal{L} \propto \prod_{i=1}^{N_{\mathrm{s}}}\left[\frac{1-P_{\mathrm{out}}}{\sqrt{2\pi\left(\sigma_{\mathrm{CBASS}}^{i}\right)^{2}}}\exp\left\{-\frac{\left(S_{\nu}^{\mathrm{CBASS},i}-b S_{\nu}^{x,i}\right)^{2}}{2\left(\sigma_{\mathrm{CBASS}}^{i}\right)^{2}}\right\} \right.\\
    \left. + \frac{P_{\mathrm{out}}}{\sqrt{2\pi\left(\sigma_{\mathrm{out}}^{2}+(\sigma_{\mathrm{CBASS}}^{i})^{2}\right)}}\exp\left\{-\frac{\left(S_{\nu}^{\mathrm{CBASS},i}-\mu_{\mathrm{out}}\right)^{2}}{2\left(\sigma_{\mathrm{out}}^{2}+\left(\sigma_{\mathrm{CBASS}}^{i}\right)^{2}\right)}\right\}\right],
\end{multline}
where $N_{\mathrm{s}}$ is the number of matched sources, $S_{\nu}^{\mathrm{CBASS},i}$ is the measured C-BASS flux density for the $i^{\mathrm{th}}$ matched source and $\sigma_{\mathrm{CBASS}}^{i}$ is the corresponding uncertainty on the C-BASS flux density. We sample our model using the No-U-Turn Sampler (NUTS), implemented in \textsc{PyMC3} \citep{JMLR:v15:hoffman14a, 2016ascl.soft10016S}. NUTS is a self-tuning variant of Hamiltonian Monte Carlo that allows for highly efficient exploration of the parameter space.

\cite{1978AJ.....83..451P} showed that approximately 60 per cent of strong radio sources at $5\,\mathrm{GHz}$ have flat spectra and are therefore likely to be compact, variable sources. Indeed, previous studies of the GB6 source catalogue have found a significant fraction of variable sources \citep{1996ApJS..103..427G,1998ASPC..144..283G,2001IAUS..205...98G}. As such, we would expect roughly half of the objects in $5\,\mathrm{GHz}$ surveys to be varying, meaning any comparison of individual flux densities observed at different epochs will be strongly contaminated by source variability. A comparison of flux densities over the whole catalogue remains useful for checking the overall consistency of the different flux density scales. However, we can use the CRATES catalogue as a proxy for potentially variable sources in the C-BASS catalogue. By matching the C-BASS catalogue with the corresponding $4.85\,\mathrm{GHz}$ sources in the CRATES catalogue, we can remove any C-BASS sources that are positively identified with a CRATES source. In doing so, we obtain a set of CRATES cleaned C-BASS sources with the flat-spectrum sources largely removed, thereby removing the primary source of variability in the C-BASS catalogue. We perform the flux-density fits described above to both the full C-BASS catalogue matches, along with the CRATES cleaned matches. The latter set of fits will be better able to detect any instrumental or systematic errors in the C-BASS flux-density scale.

The results for our fitted values of $b$, comparing to GB6 and PMN, are given in Table \ref{tab: flux ratio fit}. There is a slight discrepancy between the C-BASS flux-density scale and the PMN scale, when applying the CG30 mask and comparing to the full C-BASS catalogue. This discrepancy is removed when we remove any C-BASS sources associated with a CRATES source. We see a similar pattern for GB6, with the overall consistency being improved by cleaning the C-BASS catalogue of flat-spectrum sources. The C-BASS flux-density scale is consistent with the GB6 flux-density scale to within $\sim 4$ per cent. For the PMN catalogue there are fewer sources for comparison in the common overlap region, meaning consistency is only established in this case to $\sim 10$ per cent. Differences between the flux-density scales will result from a number of complicating factors e.g., the fact that C-BASS sources are really blends of multiple GB6/PMN sources, differing pass-bands, source variability etc. Given this, and the additional $\sim 5$ per cent uncertainties in the GB6 and PMN flux-density scales, we conclude that the C-BASS flux-density scale is consistent with these flux-density scales. We do not include any colour corrections in this analysis. The choice of $4.76\,\mathrm{GHz}$ as the effective frequency for C-BASS minimizes colour corrections to $\sim 0.02$ per cent for a source with spectral index $\alpha=0$, and $\sim 0.2$ per cent for a source with spectral index $\alpha=-1$. Colour correction for GB6/PMN are negligible due to their narrow bandwidths. Corrections due to flux-density extrapolation are also not considered here. For a typical source with spectral index, $\alpha\approx-0.83$ the correction is $\sim 1.5$ per cent. Given the uncertainties on the fitted values of $b$, and the additional uncertainties on the GB6/PMN flux-density scales, any such corrections to the flux-densities of C-BASS sources will not affect our overall conclusions here.

\begin{table}
    \centering
    \caption{Fitted values for $b$, the ratio between C-BASS and GB6 or PMN flux densities. Values were obtained by comparing flux densities between C-BASS and GB6/PMN, for both the full C-BASS catalogue and the CRATES-cleaned C-BASS catalogue. Alongside the mean fitted value for $b$, we also give the $95$ per cent highest posterior density (HPD) interval. We can see that by removing sources from the C-BASS catalogue associated with sources in the CRATES catalogue, the overall consistency between the C-BASS and GB6/PMN catalogues is improved.}
    \begin{tabular}{llcc}
        \hline
         Comparison & CRATES & $\langle b\rangle$ & 95\% $\mathrm{HPD}[b]$\\
         Catalogue & Cleaned? & & \\
         \hline
         GB6 no mask & No & 0.95 & $\{0.90, 1.00\}$\\
         GB6 CG30 mask & No & 0.95 & $\{0.91, 1.00\}$\\
         GB6 no mask & Yes & 1.04 & $\{0.97, 1.11\}$\\
         GB6 CG30 mask & Yes & 1.04 & $\{0.97, 1.11\}$\\
         PMN no mask & No & 0.95 & $\{0.89, 1.00\}$\\
         PMN CG30 mask & No & 0.82 & $\{0.68, 0.97\}$\\
         PMN no mask & Yes & 1.05 & $\{0.93, 1.17\}$\\
         PMN CG30 mask & Yes & 1.01 & $\{0.83, 1.19\}$\\
         \hline
    \end{tabular}
    \label{tab: flux ratio fit}
\end{table}

\subsection{Outlier Sources}\label{subsec: outliers}

In Fig. \ref{fig: gb6 cbass flux} we can see a number of C-BASS sources showing discrepancies with the GB6/PMN flux densities. As noted above, we expect a significant number of sources with flux densities $S_{\nu}\gtrsim 1\,\mathrm{Jy}$ to be flat-spectrum, variable sources. If we extract sources from the CRATES cleaned C-BASS catalogue that show a $\geq 3\sigma$ discrepancy between the C-BASS and GB6/PMN flux densities, we find six sources. In four of the cases the sources are buried deep within surrounding diffuse emission, with the matched GB6 flux densities being below $500\,\mathrm{Jy}$. In this situation aperture photometry does not yield reliable point estimates of individual flux densities.

We identify QSO B0723-007 as having a discrepant flux density when matching to PMN ($5.5\pm 1.2\,\mathrm{Jy}$ in C-BASS compared to $1.4\pm0.07\,\mathrm{Jy}$ in PMN). This is a BL Lacertae object, which are known to exhibit a high degree of radio variability \citep{2009AJ....137.5022N}, but is not present in the CRATES catalogue. However, it is important to note that CRATES pre-selected flat-spectrum sources for observation on the basis of spectral indices calculated between the PMN and NVSS catalogues in this region. Given the large separation in time between these surveys and the CRATES survey, we may expect CRATES to miss a small number of variable sources \citep{2007ApJS..171...61H}. The final extracted source was Virgo A. Whilst Virgo A is not a variable source, it does possess a large-scale halo that is not picked up the small beam of GB6, leading to the seemingly anomalously high value of the C-BASS flux density ($75.9\pm 1.08\,\mathrm{Jy}$). The C-BASS flux density for Virgo A is consistent with measurements in \cite{1977A&A....61...99B}. Whilst a detailed study of source variability between the C-BASS and GB6/PMN catalogues is beyond the scope of this paper, flat-spectrum sources do account for a significant degree of variability between these catalogues. This is evidenced by the improved consistency between the flux-density scales when matching to the CRATES-cleaned C-BASS catalogue, and the lack of a large additional population of variable sources beyond those identified with the CRATES catalogue.

\subsection{Differential Source Counts}\label{subsec: diff source counts}

Differential source counts have been intensively studied (e.g., \cite{2005A&A...431..893D, deZotti2010, Davies2011}), having important implications for models of cosmic evolution (e.g.,  \cite{1984ApJ...287..461C, 2010MNRAS.404..532M}) and provided the first clear evidence against the Steady State Theory \citep{1968MNRAS.139..515P}. However, for the purposes of C-BASS analysis we calculate the differential source counts to provide an additional cross-check on the statistical properties of our bright-source population.

Given that we are concerned with the statistical properties of our catalogue at flux-densities $\gtrsim 1\,\mathrm{Jy}$, where the C-BASS catalogue is largely complete, we do not consider correction factors to the differential source counts necessary at lower flux densities to account for catalogue incompleteness and varying flux-density sensitivities across the sky. Sources are put into flux-density bins of width $0.2\,\mathrm{dex}$. For each bin, source counts are determined against a central flux density, $S_{c}$, taken to be the geometric mean of flux densities in that bin. The differential source counts for a given flux-density bin are then given by
\begin{equation}
\frac{\mathrm{d}N}{\mathrm{d}S}\Bigr|_{\substack{S=S_{c}}}=n(S_{c})=\frac{N_{c}}{A\Delta S},
\label{eq: diff counts}
\end{equation}
where $N_{c}$ is the number of sources in the flux-density bin, $A$ is the sky area to which the catalogue corresponds and $\Delta S$ is the bin width. Errors are estimated for the source counts by assuming Poisson statistics. We only consider bins with $N_{c}\geq 9$ so as to maintain a $\mathrm{SNR}\geq 3$ for each bin.

\begin{figure}
\centering
\includegraphics[width=\columnwidth]{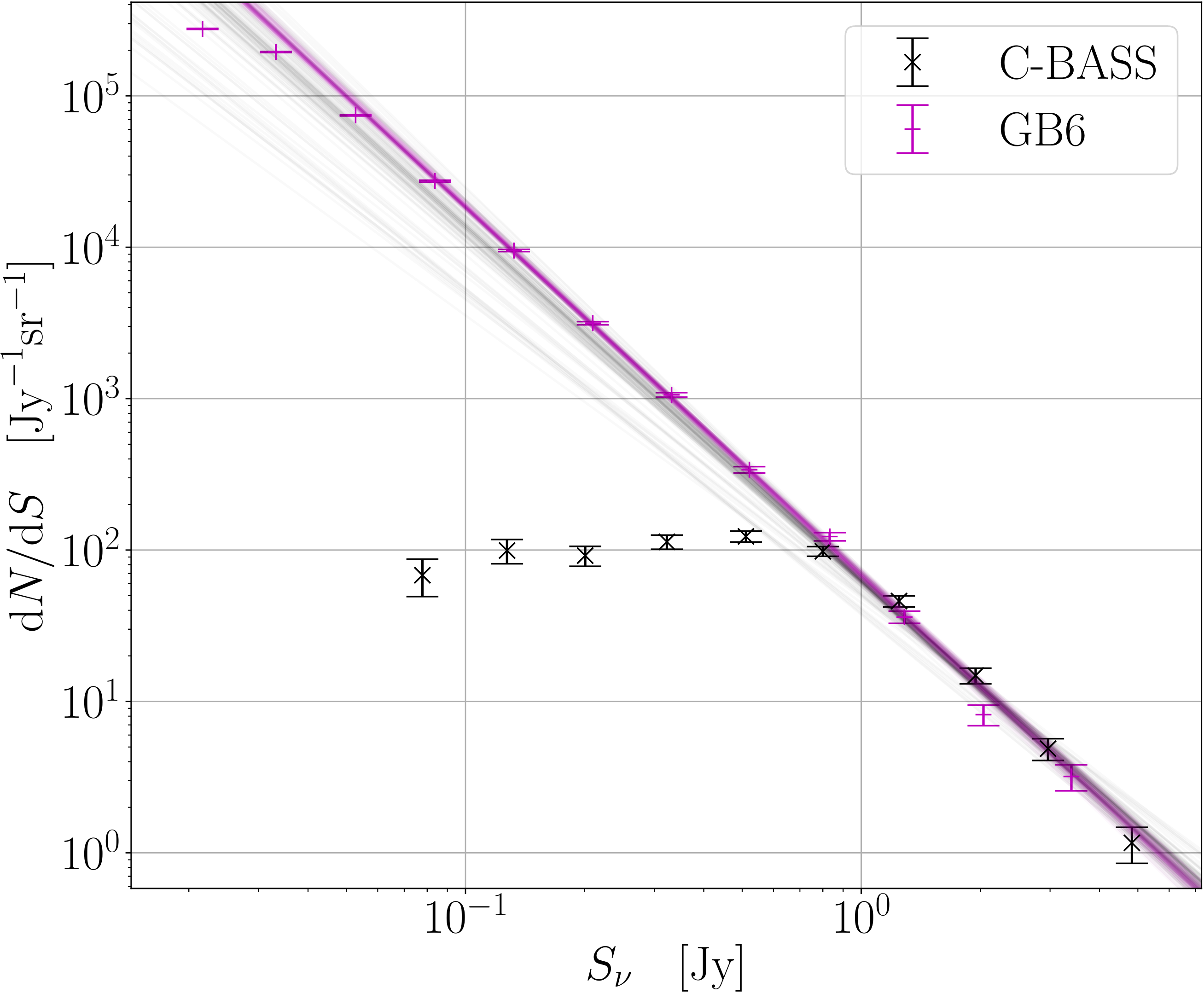}
\caption{Differential source counts calculated using the C-BASS northern sky total intensity catalogue, plotted alongside the differential source counts for the GB6 catalogue. Source counts were calculated applying the CG30 mask. We also plot as grey-scale the fitted model source counts corresponding to 100 posterior samples for the C-BASS and GB6 models. At low flux densities we note the expected lower levels in the C-BASS counts. This arises due to the low $45\,\mathrm{arcmin}$ resolution of the C-BASS map, leading to confusion issues for faint sources, along with fainter sources being more heavily obscured by diffuse emission and noise. Good agreement between GB6 and C-BASS differential source counts is found over intermediate flux densities ($\sim 1\,\mathrm{Jy}-10\,\mathrm{Jy}$).}
\label{fig:diff counts I}
\end{figure}

The differential source counts for C-BASS North in total intensity, applying the CG30 mask, are plotted in Fig. \ref{fig:diff counts I}, alongside the differential source counts for the GB6 catalogue. We can see that at low flux densities the C-BASS source counts are significantly lower. This is to be expected from the increasing catalogue incompleteness at these lower flux densities. From the differential source counts, we can again see the catalogue completeness level from the point at which the source count turn-off ends and we have good agreement between the C-BASS and GB6 counts, i.e., over flux densities $\sim 1\,\mathrm{Jy} - 10\,\mathrm{Jy}$. It is worth noting here that the errors obtained from assuming Poisson statistics are likely to be somewhat optimistic. These do not account for false positives in the source catalogues, or errors arising from incorrect binning due to uncertain flux-density estimates. These effects become significant at lower flux densities, where the proportion of false positives in the catalogue is greater, and our photometry methods become less reliable.

We fit the usual power-law model to the differential source counts, given by
\begin{equation}
\frac{\mathrm{d}N}{\mathrm{d}S} = n(S) = \Lambda \left(\frac{S}{S_{0}}\right)^{\gamma},
\end{equation}
where $\Lambda$ is the count amplitude, $S_{0}=3\,\mathrm{Jy}$ is the reference flux density and $\gamma$ is the power-law index for the source counts. For C-BASS sources we fit to data points above the turn-off in the source counts ($S_{\nu}>750\,\mathrm{Jy}$), where the C-BASS catalogue is largely complete. For GB6, we fit to data points over the range $S_{\nu}>0.07\,\mathrm{Jy}$, so as to avoid the need to apply correction factors to faint GB6 source counts. We assume a Gaussian likelihood, and assign the following weakly informative priors to $\Lambda$ and $\gamma$,
\begin{align}
    &\Lambda \sim \mathcal{N}(\mu=5, \sigma=5),\\
    &\gamma \sim \mathcal{N}(\mu=-2.5, \sigma=2.0).
\end{align}
To account for additional errors in the source counts arising from flux-density uncertainties and incorrect binning of sources, we marginalize over an additional percentage error, $f_{\mathrm{extra}}$, so that the total error on the source count in the $i^{\mathrm{th}}$ bin is given by
\begin{equation}
    \sigma_{i} = \sqrt{\sigma_{\mathrm{Poisson}}^{2} + \left(f_{\mathrm{extra}}n(S_{i})\right)^{2}},
\end{equation}
where $\sigma_{\mathrm{Poisson}}$ is the Poisson error on the source count. We assign an exponential prior to the additional percentage error, $f_{\mathrm{extra}}\sim\mathrm{Exp}(\lambda=20)$, corresponding to a characteristic value of $5$ per cent. We again explore our parameter space using the NUTS algorithm. 

The results from these power-law fits are summarized in Table \ref{tab: power-law}. When we do not apply any Galactic plane mask to the C-BASS catalogue, we fit a noticeably shallower spectral index than for the GB6 catalogue. This is driven by the excess in bright source counts in the C-BASS catalogue. This is caused by the spurious tagging of bright, highly-extended emission in the Galactic plane, and the over-estimation of source flux-densities in the presence of such emission, as discussed in Section \ref{subsec: gb6 matching}. On applying the CG30 mask, the 95 per cent HPD intervals for C-BASS and GB6 overlap as expected. This can be seen in Fig. \ref{fig:diff counts I}, where we plot 100 posterior samples from the C-BASS and GB6 models. When we apply the CG30 mask the posterior samples for the C-BASS and GB6 models overlap significantly.
\begin{table*}
\caption{Fitted power-law parameters for the differential source counts. We quote the mean parameter values and the 95 per cent HPD intervals. Results are given for cases where we apply no mask and the CG30 mask.}
\label{tab: power-law}
\begin{tabular}{lcccc}
\hline
 Catalogue & $\langle \Lambda\rangle$ & 95\% $\mathrm{HPD}\left[\Lambda\right]$ & $\langle\gamma\rangle$ & 95\% $\mathrm{HPD}\left[\gamma\right]$\\
\hline
C-BASS: & &\\
\hspace{3mm} No mask & 5.10 & \{4.43, 5.73\} & $-1.52$ & \{$-1.60$, $-1.43$\}\\
\hspace{3mm} CG30 mask & 4.73 & \{3.75, 5.70\} & $-2.37$ & \{$-2.59$, $-2.16$\}\\
GB6: & &\\
\hspace{3mm} No mask & 3.15 & \{2.75, 3.53\} & $-2.41$ & \{$-2.46$, $-2.36$\}\\
\hspace{3mm} CG30 mask & 4.55 & \{3.80, 5.28\} & $-2.45$ & \{$-2.52$, $-2.39$\}\\
\hline
\end{tabular}
\end{table*}

\section{Conclusions}\label{sec: conclusions}

We have produced a total-intensity catalogue of point sources at $4.76\,\mathrm{GHz}$, covering declinations $\delta\geq 10^{\circ}$. The catalogue was produced by filtering the whole sky using the SMHW2 filter, and contains 1784 sources. The C-BASS source catalogue has been characterized through comparisons with the pre-existing GB6 and PMN catalogues over their common declination ranges. Through these direct comparisons we estimate the C-BASS catalogue to have a $90$ per cent completeness level of approximately $610\,\mathrm{mJy}$, when applying the CG30 mask, with a corresponding reliability of $98$ per cent. The catalogue is available on request to the corresponding author, and will be made publicly available with the C-BASS northern sky data release. A sample of the first ten sources from the C-BASS catalogue is given in Appendix \ref{appndx: catalogue sample}.

These comparisons with the GB6 and PMN catalogues also enabled a number of checks to be performed on the C-BASS data. We find a modal absolute positional offset of approximately $3.2\,\mathrm{arcmin}$ when applying the CG30 mask. The low level of additional scatter in positional offsets, compared to the positional uncertainty found during the algorithm validation, acts to confirm the accuracy of the C-BASS pointing. We compare the flux densities of C-BASS sources with matched sources in the GB6 and PMN catalogues, using a measurement error model to fit for the gradient between the flux-density scales. Applying the CG30 mask, and cleaning flat-spectrum sources from C-BASS using the CRATES catalogue, we find a mean gradient of $\langle b\rangle = 1.04$ comparing to GB6, and $\langle b\rangle = 1.01$ comparing to PMN. In both cases the $95$ per cent HPD intervals overlap with $1$. Considering the $\sim 5$ per cent uncertainties in the GB6 and PMN flux-density scales, we conclude that the C-BASS and GB6/PMN flux-density scales are consistent.

As an additional check on the statistical properties of the bright source population in the C-BASS catalogue, we calculate the differential source counts for C-BASS sources. Applying the CG30 mask, the C-BASS source counts show excellent agreement with the GB6 source counts over the flux-density range $\sim 1 - 10\,\mathrm{Jy}$. At lower flux densities, we see the expected turn-off in the C-BASS source counts as the C-BASS catalogue becomes more incomplete.

\begin{figure}
    \centering
    \includegraphics[width=\columnwidth]{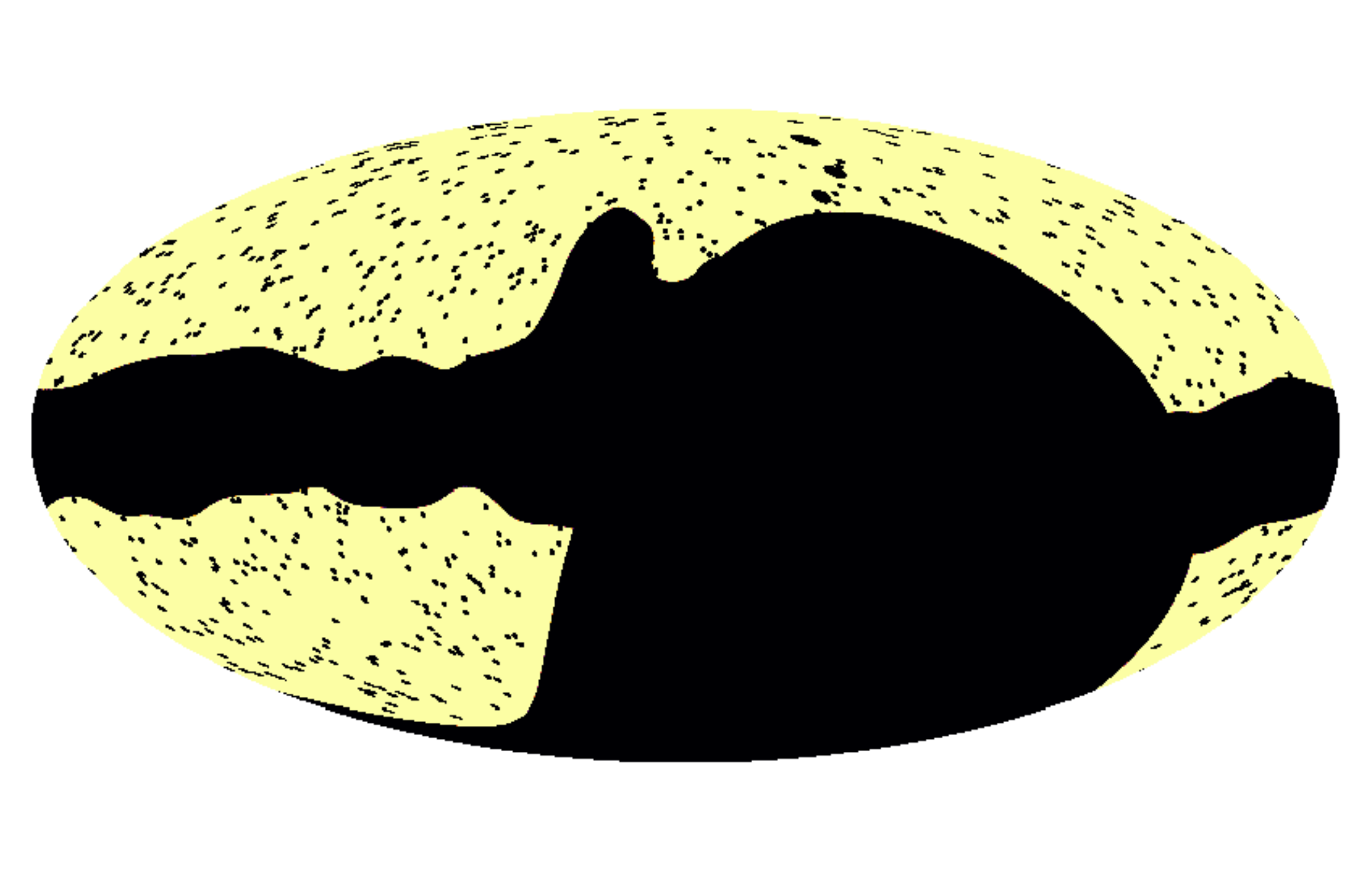}
    \caption{An example point-source mask derived from the C-BASS total intensity northern sky catalogue, shown here in Galactic coordinates with masked pixels in black. In this example we mask all sources with flux densities greater than $610\,\mathrm{mJy}$ i.e., above the 90 per cent completeness level obtained when applying the CG30 mask. We mask all pixels within $45\,\mathrm{arcmin}$ of a source, and further mask all pixels within $2^{\circ}$ of sources with flux-densities greater than $10\,\mathrm{Jy}$. In addition to masking around sources, we have masked all pixels within the CG30 mask.}
    \label{fig: source mask}
\end{figure}

For the purposes of C-BASS analysis, the primary products of this work are the checks on the data quality, and a mask of the bright point-source emission in the C-BASS map. In Fig. \ref{fig: source mask} we show an example point-source mask derived from the C-BASS catalogue, combined with the CG30 mask. Here we mask all sources with flux densities greater than $610\,\mathrm{mJy}$. We mask all pixels within $45\,\mathrm{arcmin}$ of a source, and further mask all pixels within $2^{\circ}$ of sources brighter than $10\,\mathrm{Jy}$. In future work this will be extended to include a catalogue of polarized sources. The mask produced from this will be key for the analysis of C-BASS polarization data.

\section*{Acknowledgements}

The \mbox{C-BASS} project is a collaboration between Oxford and Manchester Universities in the UK, the California Institute of Technology in the U.S.A., Rhodes University, UKZN and the South African Radio Observatory in South Africa, and the King Abdulaziz City for Science and Technology (KACST) in Saudi Arabia. It has been supported by the NSF awards AST-0607857, AST-1010024, AST-1212217, and AST-1616227, and NASA award NNX15AF06G, the University of Oxford, the Royal Society, STFC, and the other participating institutions. This research was also supported by the South African Radio Astronomy Observatory, which is a facility of the National Research Foundation, an agency of the Department of Science and Technology. 
CD and SH acknowledge support from an STFC Consolidated Grant (ST/P000649/1).
CD acknowledges support from an ERC Starting (Consolidator) Grant (no.~307209).
MWP acknowledges funding from a FAPESP Young Investigator fellowship, grant 2015/19936-1. We make use of the \textsc{Python} \textsc{matplotlib} \citep{Hunter:2007}, \textsc{numpy} \citep{numpy2006}, \textsc{scipy} \citep{scipy2001}, \textsc{healpy} \citep{2005ApJ...622..759G, Zonca2019}, \textsc{astropy} \citep{2013A&A...558A..33A, 2018AJ....156..123A}, \textsc{OpenCV} \citep{kaehler2016learning}, \textsc{PyMC3} \citep{2016ascl.soft10016S} and \textsc{theano} \citep{2016arXiv160502688full} packages.




\bibliographystyle{mnras}
\bibliography{references} 



\appendix
\section{C-BASS Catalogue Sample}\label{appndx: catalogue sample}

In Table \ref{tab: catalogue sample} we list the first ten sources from the C-BASS catalogue. Columns are arranged as,
\begin{enumerate}
    \item \textbf{SourceID}: The C-BASS source identifier.
    \item \textbf{RA [deg]}: The source right ascension in degrees (J2000 coordinates).
    \item \textbf{DEC [deg]}: The source declination in degrees (J2000 coordinates).
    \item \textbf{$l$ [deg]}: The source Galactic longitude in degrees.
    \item \textbf{$b$ [deg]}: The source Galactic latitude in degrees.
    \item \textbf{APERFLUX [Jy]}: The source flux-density estimate obtained using aperture photometry.
    \item \textbf{APERFLUX Error [Jy]}: The uncertainty on the APERFLUX estimate.
    \item \textbf{DETFLUX [Jy]}: The source flux-density estimate obtained from the peak value in the SMHW2 filtered map.
    \item \textbf{DETFLUX Error [Jy]}: The uncertainty on the DETFLUX estimate.
    \item \textbf{SNR}: The SNR at which the source was detected in the SMHW2 filtered map.
    \item \textbf{Matched}: Column denoting which catalogues contained at least one match to the C-BASS source. G denotes at least one GB6 match, P denotes at least one PMN match, R denotes at least one RATAN-600 match, and C denotes a source that was matched with a CRATES source during the GB6/PMN comparisons. A 0 means the source was not matched with any source in these ancillary catalogues. For example, GP means the source was matched with at least one GB6 source, and at least one PMN source.
    \item \textbf{CG30}: This column denotes whether the source was within the CG30 mask. A 1 means the source was inside the CG30 mask (i.e., it was masked for our analysis) and a 0 means the source was outside the CG30 mask.
\end{enumerate}

\begin{figure}
    \centering
    \includegraphics[width=\columnwidth]{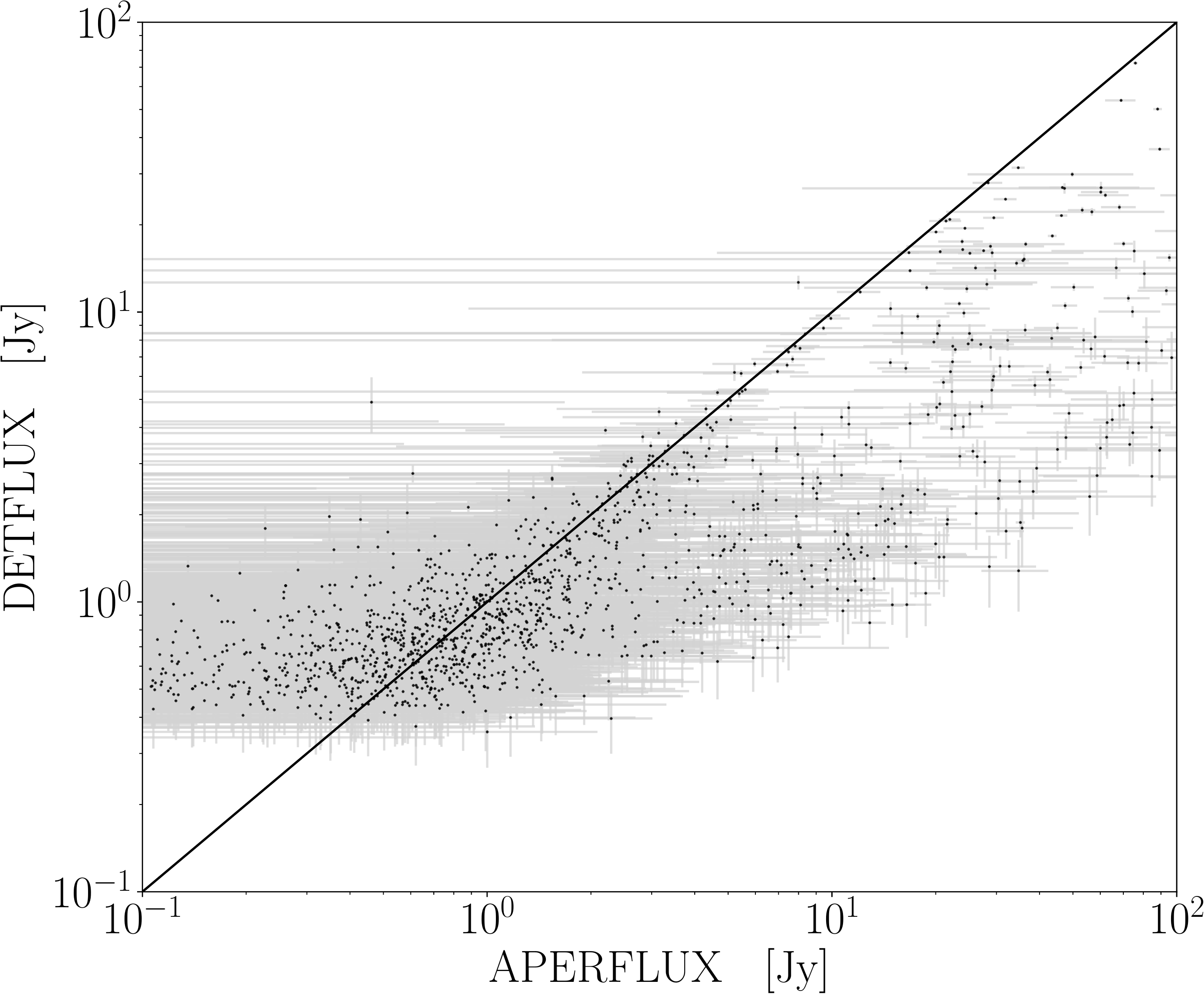}
    \caption{DETFLUX estimates plotted against APERFLUX estimates for the C-BASS catalogue. We also plot the line DETFLUX=APERFLUX. For low flux-densities ($S_{\nu}\lesssim 500\,\mathrm{mJy}$) the DETFLUX estimates plateau at the C-BASS flux-density limit. At these low flux-densities the DETFLUX results typically provide more reliable flux-density estimates than the APERFLUX results. This is because sources are detected at greater SNR in the filtered maps, from which the DETFUX peak estimates are extracted. However, for brighter sources the DETFLUX estimation procedure can miss extended emission, resulting in the flux-density estimates being systematically biased low compared to the APERFLUX estimates. This is more noticeable for sources close to the Galactic plane.}
    \label{fig: det aper}
\end{figure}

In Fig. \ref{fig: det aper} we show the DETFLUX estimates plotted against the APERFLUX estimates for the C-BASS catalogue. The DETFLUX estimates typically provide more reliable flux-density estimates for fainter sources ($S_{\nu}\lesssim 500\,\mathrm{mJy}$) and sources in regions of brighter diffuse emission. In this situation, the APERFLUX estimates become heavily contaminated by the diffuse background. This can result in the uncertainties on the APERFLUX estimates becoming larger than the APERFLUX estimates, or yield negative APERFLUX estimates due to excessive background subtraction. However, for brighter sources ($S_{\nu}\gtrsim 1\,\mathrm{Jy}$) the DETFLUX estimates are systematically biased low compared to the APERFLUX estimates. This is due to the DETFLUX estimates missing any extended emission from brighter sources. It is for this reason that we use the APERFLUX estimates in our analysis of the catalogue flux-density scales, with the DETFLUX results provided as auxiliary flux-density estimates. These issues have previously been noted in \cite{2014A&A...571A..28P,2016A&A...594A..26P}, when constructing the PCCS.

We note here that we do not provide results from any Gaussian or PSF fitting at source locations in the C-BASS map. Due to the low resolution of the C-BASS survey, and the bright diffuse emission over much of the sky, fitting for detailed PSF parameters is a major challenge for all but the brightest sources ($S_{\nu}\gtrsim 10\,\mathrm{Jy}$). For the vast majority of sources, it is extremely difficult to achieve convergence during the parameter fitting. We therefore leave this fitting to more detailed analyses of the brightest source emission.

\begin{landscape}
\begin{table}
\caption{The first ten sources from the C-BASS northern sky catalogue. In column 1 we give the C-BASS source identifier. In columns 2 and 3 we give the right ascension and declination in degrees respectively (in J2000 coordinates). In columns 4 and 5 we give the Galactic longitude and latitude in degrees respectively. In column 6 we give the APERFLUX estimate for the source flux-density in Jy. In column 7 we give the uncertainty on the APERFLUX estimate in Jy. In column 8 we give the DETFLUX estimate in Jy, determined directly from the peak value of the source in the SMHW2 filtered map. In column 9 we give the uncertainty on the DETFLUX estimate in Jy. In column 10 we give the detected SNR of the source in the filtered map. This is equivalent to the DETFLUX estimate divided by its corresponding uncertainty. In column 11 we denote whether the C-BASS source was matched with sources in the GB6, PMN, RATAN-600 or CRATES catalogues. G means a source was matched with at least one GB6 source, P means a source was matched with at least one PMN source, R means a source was matched with at least one RATAN-600 source, and C means a source was matched with at least one CRATES source during the GB6/PMN comparisons. A 0 means the C-BASS source was not matched with any source in these ancillary catalogues. In column 12 we denote whether the C-BASS source was inside the CG30 mask. A 1 means the source was inside the CG30 mask and a 0 means the source was outside the CG30 mask.}
\label{tab: catalogue sample}
\begin{tabular}{llllllllllll}
\toprule
             SourceID & RA [deg]& DEC [deg] & $l$ [deg] & $b$ [deg] & APERFLUX & APERFLUX & DETFLUX & DETFLUX &    SNR & Matches & CG30 \\
             & (J2000) & (J2000) & & & [Jy] & Error [Jy] & [Jy] & Error [Jy] & & &\\
\midrule
 CBASS\_J000056+553448 &             0.23 &            +55.58 &  115.79 &   +6.59 &          0.43 &                5.23 &         0.72 &               0.19 &    3.8 &       G &    1 \\
 CBASS\_J000110+405055 &             0.29 &            +40.85 &  112.78 &  +21.03 &          0.11 &                0.93 &         0.66 &               0.13 &    5.2 &       G &    0 \\
 CBASS\_J000137+643717 &             0.40 &            +64.62 &  117.62 &   +2.26 &          6.30 &                3.84 &         2.41 &               0.44 &    5.4 &       0 &    1 \\
 CBASS\_J000249+671313 &             0.71 &            +67.22 &  118.24 &   +4.79 &        127.65 &                4.29 &        40.92 &               0.31 &  133.2 &       0 &    1 \\
 CBASS\_J000316+655840 &             0.82 &            +65.98 &  118.05 &   +3.56 &         32.00 &                4.03 &         1.75 &               0.35 &    5.1 &       0 &    1 \\
 CBASS\_J000433+125050 &             1.14 &            +12.85 &  105.56 &  +48.44 &          1.69 &                1.20 &         1.04 &               0.14 &    7.3 &       G &    0 \\
 CBASS\_J000548+683650 &             1.45 &            +68.61 &  118.77 &   +6.11 &         42.21 &                4.16 &         6.21 &               0.26 &   24.1 &       0 &    1 \\
 CBASS\_J000609+381223 &             1.54 &            +38.21 &  113.23 &  +23.81 &          1.11 &                0.93 &         0.86 &               0.14 &    6.0 &      GC &    0 \\
 CBASS\_J000618+722000 &             1.58 &            +72.33 &  119.48 &   +9.76 &          9.26 &                2.46 &         2.56 &               0.17 &   14.8 &       G &    1 \\
 CBASS\_J000622-062113 &             1.59 &             -6.35 &   93.63 &  -66.63 &          1.77 &                1.06 &         2.19 &               0.13 &   16.8 &      PC &    0 \\
\bottomrule
\end{tabular}
\end{table}
\end{landscape}


\bsp	
\label{lastpage}
\end{document}